\documentclass[11pt]{article}
\usepackage{amsfonts,amssymb, graphicx,a4}

\newtheorem{theo}{Theorem}[section]
\newtheorem{lem}[theo]{Lemma}
\newtheorem{prop}[theo]{Proposition}
\newtheorem{defi}[theo]{Definition}
\newtheorem{coro}[theo]{Corollary}
\newtheorem{rema}[theo]{Remark}

\begin{document}


\let\a=\alpha \let\b=\beta
\let\d=\delta \let\e=\varepsilon
\let\f=\varphi \let\g=\gamma \let\h=\eta    \let\k=\kappa \let\l=\lambda
\let\m=\mu \let\n=\nu \let\om=\omega    \let\p=\pi \let\ph=\varphi
\let\r=\rho \let\s=\sigma \let\t=\tau \let\th=\vartheta
\let\y=\upsilon \let\x=\xi \let\z=\zeta
\let\D=\Delta \let\F=\Phi \let\G=\Gamma \let\L=\Lambda \let\Th=\Theta
\let\O=\Omega \let\P=\Pi \let\Ps=\Psi \let\Si=\Sigma \let\X=\Xi
\let\Y=\Upsilon

\global\newcount\numsec\global\newcount\numfor
\gdef\profonditastruttura{\dp\strutbox}
\def\senondefinito#1{\expandafter\ifx\csname#1\endcsname\relax}
\def\SIA #1,#2,#3 {\senondefinito{#1#2}
\expandafter\xdef\csname #1#2\endcsname{#3} \else
\write16{???? il simbolo #2 e' gia' stato definito !!!!} \fi}
\def\etichetta(#1){(\veroparagrafo.\veraformula)
\SIA e,#1,(\veroparagrafo.\veraformula)
 \global\advance\numfor by 1
 \write16{ EQ \equ(#1) ha simbolo #1 }}
\def\etichettaa(#1){(A\veroparagrafo.\veraformula)
 \SIA e,#1,(A\veroparagrafo.\veraformula)
 \global\advance\numfor by 1\write16{ EQ \equ(#1) ha simbolo #1 }}
\def\BOZZA{\def\alato(##1){
 {\vtop to \profonditastruttura{\baselineskip
 \profonditastruttura\vss
 \rlap{\kern-\hsize\kern-1.2truecm{$\scriptstyle##1$}}}}}}
\def\alato(#1){}
\def\veroparagrafo{\number\numsec}\def\veraformula{\number\numfor}
\def\Eq(#1){\eqno{\etichetta(#1)\alato(#1)}}
\def\eq(#1){\etichetta(#1)\alato(#1)}
\def\Eqa(#1){\eqno{\etichettaa(#1)\alato(#1)}}
\def\eqa(#1){\etichettaa(#1)\alato(#1)}
\def\equ(#1){\senondefinito{e#1}$\clubsuit$#1\else\csname e#1\endcsname\fi}
\let\EQ=\Eq
\def\GI{\mathbb{G}}
\def\VU{\mathbb{V}}
\def\vv{\vskip.2cm}
\def\vvv{\vskip.3cm}
\def\v{\vskip.1cm}


\def\sqr#1#2{{\vcenter{\vbox{\hrule height.#2pt
\hbox{\vrule width.#2pt height#1pt \kern#1pt
\vrule width.#2pt}\hrule height.#2pt}}}}
\def\square{\mathchoice\sqr34\sqr34\sqr{2.1}3\sqr{1.5}3}

\def\\{\noindent}
\def\EE{\mathbb{E}}
\def\Z{\mathbb{Z}}
\def\GG{\mathcal{G}}
\def\QQ{\mathcal{Q}}
\def\TT{\mathcal{T}}
\def\AA{\mathcal{A}}
\def\BB{\mathcal{B}}
\def\FF{\mathcal{F}}
\def\PP{\mathcal{P}}
\def\RR{\mathcal{R}}
\def\SS{\mathcal{S}}
\def\ES{\mathbf{S}}
\def\EP{\mathbf{P}}
\def\LL{\mathcal{L}}
\def\0{\emptyset}
\def\N{\mathbb{N}}
\def\setn{{\rm I}_n}
\def\CC{\mathcal{C}}
\def\E{\mathcal{E}}
\def\ER{{\bf R}}
\def\Lad{{\mathbb{L}^d}}
\def\Ed{{\mathbb{E}^d}}
\def\Zd{\mathbb{Z}^d}
\def\supp{{\rm supp}\,}

\def\arm{{}}
\font\bigfnt=cmbx10 scaled\magstep1


\title{Convergent expansions \\
for Random Cluster Model with $q>0$ on infinite graphs }
\author{
Aldo Procacci\\
\small{Departamento de Matem\'atica UFMG}\\
\small{ 30161-970 - Belo Horizonte - MG
Brazil}\\
\\
Benedetto Scoppola \\
\small{ Dipartimento di Matematica
 Universit\'a ``Tor Vergata'' di Roma}\\
\small{ V.le della ricerca scientifica,
00100 Roma, Italy}
}
\maketitle

\begin{abstract}
In this paper we extend our previous results on the connectivity functions
and pressure of the Random Cluster Model in the
highly subcritical phase and in the highly supercritical phase,
originally proved only on the cubic lattice $\Z^d$,
to a much wider class of infinite graphs.
In particular, concerning the subcritical regime, we show that  the connectivity functions
are analytic and decay exponentially in any bounded degree graph.
In the supercritical phase, we are able to prove the analyticity of  finite
connectivity functions in a smaller class of graphs, namely,
bounded degree graphs with the so called minimal cut-set property
and satisfying a (very mild) isoperimetric inequality. On the other  hand we  show that the
large distances decay of finite connectivity in the supercritical regime
can be polynomially slow depending on the
topological structure of the graph.
Analogous analyticity results are obtained for the pressure of the Random Cluster Model on an
infinite graph, but with the further assumptions
of amenability and quasi-transitivity of the graph.
\end{abstract}

\numsec=1\numfor=1

\vskip.5cm
\section{Introduction}

\\In recent years there has been an increasing interest about
statistical mechanics systems and stochastic processes on general infinite
graphs. The main motivation has been  the possible connections and   applications in computer science,
with particular attention to reliability of large network (e.g.  the internet). More recently, see e.g. \cite{So,So2, PSG,Bo, FP2},
people started to realize that ideas and methods of statistical mechanics could be useful to answer
questions arising in combinatorics and graph theory.

Rigorous results on this subject have appeared since the early
nineties and nowadays there is a consistent literature on this
subject. Actually, the study of  statistical mechanics and
percolation processes on infinite graphs other than the usual unit
cubic lattice $\mathbb{Z}^d$ or planar triangular and hexagonal
lattices has been limited essentially to non amenable graphs.
Roughly speaking, the non amenable graphs are those for which the
ratio of the boundary of the graph and its interior does not go to
zero in the infinite volume limit, while for amenable graphs this
ratio goes to zero. Within the class of non amenable graphs, the
study has been mostly focused on trees (i.e., graphs with no
circuits) , see e.g. \cite{BS1,BS2,J,H,HJL1,HJL2,Sh, L1,L2,L3,
GN,GJ}. There have been also a few papers dealing with percolation
processes on quasi transitive or transitive graphs, including
amenable graphs, see e.g. \cite{BB}, \cite{BLPS}, \cite{MP}.
Roughly speaking, in a transitive graph $G$  any vertex of the
graph is equivalent; in other words $G$ ``looks the same'' by
observers sitting in different vertices. In a quasi-transitive
graph $G$ there is a finite number of different types of vertices
and $G$ ``looks the same'' by observers sitting in vertices of the
same type.

Some general
results about percolation on
general infinite graphs (i.e. not necessarily non amenable and/or quasi-transitive)
appeared in
\cite{BS1}, \cite{BS2}, and \cite{BB}, and more recently in \cite {PS7}
(see also references therein). There are
also some other works about the Potts model, in particular
the antiferromagnetic case, on general  finite graphs \cite{So,So2} and on
amenable quasi-transitive infinite graphs \cite{PSG}.

\\In this paper we focus our attention on the study of the dependent
percolation process known as Random Cluster Model (RCM) on general
infinite graphs.

\\The RCM was
proposed by Fortuin and Kasteleyn in the early seventies \cite{FK} as
a generalization of
the Potts model.
The RCM on a graph $G$ depends on two real parameters: the parameter
$p\in [0,1]$ and the parameter
$q\in (0,+\infty)$. The parameter $p$ represents the weight of an
edge of $G$ to be open
independently of the other edges and it is related to the
temperature of the Potts model.
The parameter $q$, when different from 1,
introduces a dependence in the percolation process described
by RCM, and,  when integer greater than 2, it
represents the number of colors in the Potts model.

Some results on RCM can be proved for all the
values of the parameters $q$ and $p$. In particular,  there are
results about the logarithm of the total weight of the measure (pressure). Namely,
the existence of the pressure of the RCM, its independency on
boundary conditions and its differentiability
(with respect to $p$ almost everywhere in the interval $[0,1]$)
have been proved for all $q\in (0,\infty)$
when the underlying infinite graph is the  cubic lattice $\Z^d$ in
\cite{Gr}  (see also \cite{J} for some generalization of such results
to transitive amenable graphs).
This
shows that in these cases the whole machinery of the statistical mechanics, and
its probabilistic counterpart, can be used for all the values of
the parameters of the RCM.

However, the study of the statistical mechanics properties of RCM
has been developed
so far mainly in $\Z^d$, and only in the region $q\ge 1$, where
the powerful tool given by the so-called FKG inequalities is availabe.
In particular,  by
comparison inequalities (see \cite{FK}, \cite{A} and \cite{G2}), it is possible to prove that
the RCM on $\Z^d$ admits, for $q\ge 1$, a (non trivial) critical value $p_c(q)\in (0,1)$ such that for $p<p_c(q)$ the
probability to have an infinite open cluster is zero, while for
$p>p_c(q)$ is one (\cite{A}, Theorem 4.2). Many other important results can be collected for the RCM  on $\Z^d$ in the regime $q\ge 1$.
We
refer the reader to the monograph  \cite{G2} and book \cite{G3} for a detailed description of these results and references.

Concerning the case $q<1$, due to the lack of validity of  FKG inequalities  in this regime,
nearly quoting the words of Grimmett in \cite{G2}, many fundamental questions are unanswered to date,
and the theory of RCM remains obscure when $q<1$.
We  tried to answer to some of these questions
in  a recent paper \cite{PS8}, where we studied, by mean of cluster expansion methods,  the statistical mechanics behavior of the Random Cluster Model
on the cubic lattice $\Z^d$ ($d\ge 2$) for $p$ near either 0 or 1
and for all $q>0$, proving the analyticity
of the pressure and of finite connectivities in both regimes. The results of \cite{PS8} also give a
generalization  of theorem 4.2. in \cite{A}
for values of $q$ in the interval $0<q<1$.

In the present paper, by taking advantage of the robustness and malleability of cluster expansion methods, we
continue the analysis of the statistical mechanics behavior of the RCM, and in particular its analyticity
properties for $p$ near either 0 or 1, extending the results of  \cite{PS8}
to RCM on a class of graph much more general than the regular lattices like $\Z^d$.
Here we are motivated by recent results
\cite{So}, \cite{So2} \cite{PSG}, \cite{PS7} showing how statistical mechanics techniques (and in particular cluster expansion)
may give interesting contributions to specific problems concerning graph theory.

Our results are stated  in a detailed form  in theorems \ref{sub}, \ref{presub}, \ref{sup}, \ref{presup}.
However, for the benefit of the readers, we report sketchily these results here below.

For the subcritical regime  we obtain that, for any fixed  value of $q>0$ there is  $R^{sub}_q>0$ such that
for any $p$ in the disk $|p|\le R^{sub}_q$  we have the following
results:
\vskip.1cm
\\{\bf 1a)} The $n$-point
connectivity functions ($n\in \N$) of the RCM on an infinite
graph $\GI$ can be written explicitly
as analytic functions of $p$ whenever
$\GI$ is bounded degree and they decay exponentially fast at large distances,
which also implies that the probability to
have an infinite open cluster in the graph is zero when $p\in [0, R^{sub}_q)$.
These results have been obtained via a limit procedure on
sequences of subgraphs of $\GI$, and we are able to prove that the limit of the
$n$ point connectivity function
tends to the same analytic function  for free and wired boundary condition.

\vskip.1cm
\\{\bf 1b)} The pressure is analytic  in $p$ in the same region whenever $\GI$ is
quasi-transitive  and amenable.
\vskip.2cm
For the supercritical  regime  we obtain that, for any fixed  value of $q>0$ there exists
$R^{sup}_q>0$ such that
for any $p$ in the disk $|1-p|<R^{sup}_q$  we have the following
results:

\vskip.1cm
\\{\bf 2a)} For any $n\ge 1$, the $n$-point
finite connectivity function of the RCM on an infinite
graph $\GI$ can be written explicitly
as an analytic function of $1-p$  whenever $\GI$ is
bounded degree and satisfies some additional properties,
including a very weak isoperimetric inequality (see below). Such result immediately implies
that for any $p$ in the interval $(1-R^{sup}_q,1]$ the probability to have an infinite open cluster in the graph containing a fixed vertex
is strictly greater than zero.

We remark that the  class of graphs for which we can prove analyticity of correlations in the supercritical regime is
smaller than the class of bounded degree graphs, but it is still very large class: e.g.,  it
contains  $\Z^d$ and all the regular lattices
and also graphs without symmetries.
This result is
obtained with a limit construction on finite  subgraphs of $\GI$, independently
of free or wired boundary conditions. Differently from the subcritical regime,
the finite connectivity functions may decay in the supercritical phase with a rate that can be
polynomially slow, depending on the topological structure of the graph.
We plan to investigate in
details this feature of the supercritical phase on general graphs
in a forthcoming paper devoted only to Bernoulli percolation
(i.e. Random Cluster Model with $q=1$). Indeed results of this paper suggest
that the decay rate of  finite connectivities for the Bernoulli percolation
process on an infinite graph
can be adopted as an efficient and quantitative measure of the degree of connection of the graph. Namely,
the more rapid is the decay rate of connectivities, the more dense (or connected) is the graph.

\vskip.1cm
\\{\bf 2b)} The pressure is analytic  in $1-p$ if $\GI$ is in the class above and it is
(vertex and edge) quasi-transitive and amenable.

 \vskip.1cm
Our  conditions on the structure of the graph guaranteeing the convergence of the cluster expansion in the subcritical phase are quite general.
In particular, for the  existence and convergence of the connectivity functions,
it is just required for the graph to be bounded degree, which constitutes a very large class of graphs.
However, it is possible that with similar techniques one can study unbounded degree graphs in which
the vertices with large degree are  "rare enough''.
The requirement of amenability and quasi-transitivity
for the existence of the pressure is also largely expected. Roughly speaking, amenability guarantees the possibility to perform the thermodynamic limit
in the Van-Hove sense, so that the effects of the boundary vanish in the infinite volume limit.
Quasi-transitivity plays the role of "translational invariance" in the graph which is
in general a necessary tool for the existence of the pressure.

On the other hand, in the supercritical case we think that the conditions above are far from optimal. In particular, the  isoperimetric condition  is due
to technical reasons in view to adapt  the Peierls argument and contour theory to general graphs.

\vskip.1cm

The  paper is organized as follows. In section 2 we give some
definitions about  graphs. In  section 3
we introduce the model, first on finite graphs and then on infinite graphs.
In section 4 we study the highly subcritical phase,
and state
two theorems (theorem \ref{sub} and theorem \ref{presub}),
the first one concerning
the connectivity functions and the second one concerning the pressure.
The rest
of the section is devoted to the proof of these two theorems.
In section 5 we perform the analysis of the supercritical phase.
Namely, in subsection 5.1 we give some
more definitions and properties about cut sets in infinite graphs and,
at the end of the subsection,
we state the results on
the supercritical phase in form of two more theorems: theorem \ref{sup} concerns the connectivity functions
and theorem \ref{presup} concerns  the pressure.
In section 5.2
we construct the polymer expansion for the connectivity functions.
In section 5.3 we show that
this expansion is absolutely convergent for $p$ sufficiently close to 1 and
we conclude the proof of theorem \ref{sup}.
In  section 5.4
we prove theorem \ref{presup}.

\numsec=2\numfor=1

\section{Some definitions about graphs}

\\For any finite or countable set $V$, we will denote by $|V|$
the cardinality of $V$.
We denote
by ${\rm P}_n(V)$ the set of all subsets $U\subset V$ such
that $|U|=n$ and we denote by ${\rm P}_{\ge n}(V)$ the
set of all {\it finite} subsets $U\subset V$ such
that $n\le |U|<+\infty$. A {\it graph} is a pair $G=(V, E)$ with
$V$ being  a countable set, and  $E\subset {\rm P}_2(V)$. The elements of $V$
are called {\it vertices} of $G$ and the elements of $E$ are called
{\it edges} of $G$. A graph $G=(V,E)$ is {\it finite} if $|V|<\infty$, and infinite otherwise.
Let $G=(V,E)$ and $G'=(V',E')$ be two graphs. Then $G\cup G'= (V\cup V', E\cup E')$.
If $V'\subseteq V$
and $E'\subseteq E$, then $G'$ is a {\it subgraph} of $G$, written as $G'\subseteq G$.

\\Two vertices $x$ and $y$ of $G$ are {\it adjacent}
if $\{x, y\}$ is an edge of $G$.
The {\it degree} $d_x$ of a vertex $x\in V$
in $G$ is the number of vertices $y$ adjacent to $x$.
A graph $G=(V,E)$ is
{\it locally finite}
if $d_x<+\infty$ for all $x\in V$, and it is {\it bounded degree, with maximum degree $\D$,} if
$\max_{x\in V} \{d_x\}\le \D<\infty$.
A graph $G=(V, E)$
is {\it connected}
if for any pair $B, C$ of  subsets of $V$ such that
$B\cup C =V$ and $B\cap C =\emptyset$, there is an edge  $e\in E$ such
that $e\cap B\neq\emptyset$ and $e\cap C\neq\emptyset$.
A graph $G=(V,E)$ is a called a {\it tree graph} or simply a {\it tree} if it is
connected and $|E|=|V|-1$.
\def\sg {{\rm supp} \,g}
\def\sG {{\rm supp} \,G}

\\Hereafter  the symbol $\GI=(\VU,\EE)$ will denote an {\it infinite and connected} graph.

\\A {\it path} in a graph $G$ is a sub-graph $\t=(V_\t,E_\t)$ of $G$ such that
$$
V_\t=\{x_1, x_2, \dots , x_n\}~~~~
E_\t=\left\{\{x_1, x_2\}, \{x_2,x_3\},\dots , \{x_{n-1},x_n\}\right\}
$$
where all $x_i$ are distinct. The vertices $x_1$ and $x_n$ are
called end-vertices of the path, while
the vertices $x_2, \dots, x_{n-1}$ are called the inner vertices of $\t$ and we say that
$\t$ connects (or links) $x_1$ to $x_n$, (as well as $\t$ is a path from $x_1$ to $x_n$).
The length $|\t|$ of a path $\t=(V_\t,E_\t)$ is the number of its edges, i.e. $|\t|=|E_\t|$.
A path in $G$ is also called  a {\it self avoiding walk (SAW)} in $G$.

\\Given a graph $G=(V,E)$ and two distinct vertices $x,y\in V$,
we denote by ${\cal P}^{xy}_G$ the set of all paths in $G$ connecting $x$ to
$y$. The {\it distance} $d_G(x,y)$ between   two vertices $x,y$
of $G$ is the number $d_G(x,y)= \min\{|\t|: \t\in {\cal P}^{xy}_G\}$.
Note that $d_G(x,y)=1$ if and only if $\{x,y\}\in E$.
Given two edges $e$ and $e'$ of  $G$, we define
$d_G(e,e')= \min\{d_G(x,y): x\in e, y\in e'\}$.
If $S,R\subset V$ then
$d_G(S,R)=\min\{d_G(x,y): x\in S, y\in R\}$. If $F,H\subset E$ then
$d_G(F,H)= \min\{d_G(e,e'): e\in F, e'\in H\}$.

\\Let $\GI=(\VU,\EE)$  be an infinite connected graph.
A {\it ray} $\r=(\VU_\r, \EE_\r)$ in $\GI$  is an {\it infinite} sub-graph of $\GI$ such that

$$
\VU_\r=\{x_0,x_1, x_2, \dots , x_n, \dots\}~~~~
\EE_\r=\left\{\{x_0, x_1\}, \{x_1, x_2\}, \{x_2,x_3\},\dots , \{x_{n-1},x_n\},\dots\right\}
$$
where all $x_i$ are distinct. The vertex $x_0$ is called the {\it starting vertex} of the ray
and we say that $\r$ {\it starts at $x_0$}.
We denote by ${\cal R}^{x}_\GI$ the set of all rays in $\GI$ starting at  $x$.
A  ray $\r=(\VU_\r, \EE_\r)$ in $\GI$ with starting vertex $x_0$ is  {\it geodesic }   if
$d_G(x_0, x_n)=n$ for all $x_n\in \VU_\r$.

\\Let $\r$ and $\r'$ be two geodesic rays with the same  starting vertex  $x$
with vertex sets $\VU_\r=\{x_0=x, x_1, x_2, \dots , x_n,\dots\}$
and $\VU_{\r'}=\{y_0=x, y_1, y_2, \dots , y_n,\dots\}$ respectively. If
$\VU_\r$ and $\VU_{\r'}$ are such that
$d_G(x_n,y_m)=n+m$ for any $\{n,m\}\in \mathbb{N}$,
then the union $\d=\r\cup \r'$ is called
a {\it geodesic diameter} (or {\it bi-infinite geodesic}) in $\GI$.

\\Given $G=(V, E)$ connected and $R\subset V$,
let $E|_R=\{\{x,y\}\in E: x\in R, y\in R\}$
and define the graph $G|_{R}=(R, E|_R)$.
Note that $G|_{R}$ is a sub-graph of $G$.
We call $G|_{R}$ {\it the restriction of $G$ to $R$}. We say
that  $R\subset V$ is {\it  connected}
if $G|_{R}$ is connected.
Analogously, Given $G=(V, E)$ connected and $\eta \subset E$, let
$V|_\eta=\{x\in  V: x\in e ~{\rm for~some ~}e\in \eta\}$. We call $V|_\eta$ the
{\it support} of $\eta$. We say that a edge set $\eta\in E$ is {\it connected}
if the graph $g=(V|_\eta, \eta)$ is connected.

\\For any non empty $R\subset V$, we denote by
$\partial_e R$ the (edges) boundary of $R$ defined by
$$
\partial_e R = \{e\in E-E|_R: |e\cap R|=1 \}
\Eq(bounda)
$$

\\We also denote by
$\partial^{\rm ext}_v R$ the {\it  external vertex boundary}  of $R$ the subset of
$V\backslash R$
given by
$$
\partial^{\rm ext}_v R =\{v\in V\backslash R:~~  \exists e\in  E: e=\{v,v'\}~~{\rm with}~ ~ v'\in  V\}
\Eq(boundav)
$$
and we denote by
$\partial^{\rm int}_v R$ the {\it  internal vertex boundary}  of $R$ the subset of
$R$
given by
$$
\partial^{\rm int}_v R =\{v\in  R:~~  \exists e\in  E: e=\{v,v'\}~~{\rm with}~ ~ v'\in  V\backslash R\}
\Eq(boundav1)
$$
If $R\subset V$ we denote
$$
{\rm diam}(R)=\sup_{x,y\in R}d_G(x,y)\Eq(rayW)
$$
and call it the {\it diameter} of $R$.

\\Let $g=(V_g,E_g)$ be a subset of $G$ then we define $\partial g$ the (edge) boundary of $g$
as
$$
\partial g=\{e\in E-E_g: ~e\cap V_g\neq \emptyset\}
$$

\\Note that $\partial( G|_R)=\partial_e R$.

\\Let $G=(V,E)$ be a graph and let $x\in V$ and $R>0$. We denote by $B(x, R)$ the ball of radius $R$ and center at $x$,
namely $B(x, R)=\{y\in V:~d_G(x,y)\le R\} $.

\begin{defi}\label{treed}.
Let $\GI=(\VU,\EE)$ be an infinite connected graph and let $X\subset \VU$ finite. Let now $\TT_X$ denote
the set of all trees with vertex set $X$ (we recall that a tree in $X$ is a connected graph $\t=(V_\t,E_\t)$ with
$V_\t=X$ and $|E_\t|=|X|-1$).
We define
the minimal tree distance  $d_\GI^{\rm tree}(X)$  of $X$ in $\GI$, as
$$
d_\GI^{\rm tree}(X) = \min_{\t\in \TT_X}\sum_{\{x,y\}\in E_\t}
d_\GI(x,y)\Eq(treedist)
$$
\end{defi}
We remark that in this definition $X$ is not necessarily connected in $\EE$.
So $E_\t$ is not necessarily a subset of $\EE|_X$, so the pair $\{x,y\}$ does not, in general,
belong to $\EE$, and for that pair $d_\GI(x,y)>1$.
On the other hand, note that when $X$ is connected in $\GI$ then it is always possible to find some tree $\t$ in $\TT_X$ such that
$d_\GI(x,y)=1$ for any  pair $\{x,y\}\in \t$ and hence in this case $d_\GI^{\rm tree}(X)=|X|-1$.

\begin{defi}\label{connective constant}.
Let $\GI=(\VU,\EE)$ be a connected and  infinite graph. We define
the connective constant $C_\GI$, of $\GI$  as
$$
C_\GI= \sup_{n\in \N}\sup_{x\in \VU}[c_x(n)]^{1/n}\Eq(treedist2)
$$
with  $c_x(n)$ being the number of all paths (i.e. Self Avoiding Walks) of length $n$ with starting point $x$. By definition,
for any infinite graph $\GI$, we have that  $C_\GI\in [0,+\infty)\cup\{+\infty\}$.

\end{defi}
For example, for a regular tree $\mathbb{T}_k$ of degree $k$, $C_{\mathbb{T}_k}=k$. For $\Z^2$ the connectivity constant
is not known exactly but it is known to belongs to the interval $[2,62,2,68]$

\\ An {\it automorphism} of a graph $G=(V,E)$ is a bijective map
$\g:V\to V$ such that $\{x,y\}\in E \Rightarrow \{\g x, \g y\}\in E$.
A graph $G=(V, E)$ is called {\it transitive} if, for any $x, y\in V$,
there exists an automorphism $\g$ of $G$ such that  $\g(x)=y$.

An infinite connected graph $\GI=(\VU,\EE)$
is called {\it vertex  quasi-transitive} ({\it edge  quasi-transitive}) if $\VU$ ($\EE$) can be partitioned in finitely many sets
$\mathbb{O}_1, \dots \mathbb{O}_s$ (orbits) such that for $\{x, y\}\in \mathbb{O}_i$ ($\{e, e'\}\in \mathbb{O}_i$)
it exists an automorphism $\g$ on $\mathbb{G}$ which maps $x$ to $y$ ($e$ to $e'$) and this
holds
for all $i=1, \dots , s$. If $x\in \mathbb{O}_i$ and $y\in \mathbb{O}_i$ ($e\in \mathbb{O}_i$ and $e'\in \mathbb{O}_i$ ) we say that
$x$ and $y$ ($x$ and $y$) are equivalent.

Roughly speaking in a transitive infinite graph  any
vertex of the graph is equivalent; in other words
$\GI$ ``looks the same'' by observers sitting in different vertices.
In a quasi-transitive infinite graph there is
a finite number of different type of vertices and $\GI$ ``looks the same''
by observers sitting in vertices of the same type.

 As an immediate example all periodic
lattices with the elementary cell made by one site (e.g. square lattice,
triangular lattice, hexagonal lattice, etc.) are transitive infinite graphs, while
periodic
lattices with the elementary cell made by more than one site are
quasi-transitive infinite graphs.

\begin{defi}
Let $\mathbb{G}=(\mathbb{V}, \mathbb{E})$ be a connected infinite graph.
$\mathbb{G}$ is said to be {\it amenable} if
$$
\inf\left\{{|\partial_e W|\over|W|}: W\subset \mathbb{V}, ~0<|W|<+\infty\right\}=0
$$
A sequence $\{V_N\}_{N\in \mathbb{N}}$  of
finite sub-sets of $\mathbb{V}$ in an amenable graph $\GI=(\VU,\EE)$ is called a {\it F\o lner sequence} if
$$
\lim_{N\to \infty}{ |\partial_e V_N|\over|V_N|}=0
\Eq(Folner)
$$
\end{defi}
Note that such definition reminds the notion of Van Hove sequence
in statistical mechanics.
\begin{defi}
\label{folner}
Let $\VU$ be an infinite countable set. We say that a sequence  $\{V_N\}_{N\in \mathbb{N}}$  of
$\mathbb{V}$ {\it tends monotonically to $\mathbb{V}$}, and we write $V_N\nearrow\mathbb{V}$,  if,
for all $N\in \mathbb{N}$, $V_N$
is connected,  $V_{N}\subset
V_{N+1}$, and $\cup_{N\in \mathbb{N}}V_N=\mathbb{V}$.
\end{defi}
Roughly speaking, amenability in an infinite connected graph $\GI=(\VU,\EE)$ means  that the boundary of
finite connected set $X\subset \VU$ grows slower than its interior as soon as $X\nearrow\VU$.
For example, $\Z^d$ is amenable, while the regular tree $\mathbb{T}_k$  for $k\ge 3$ is not amenable.

\vskip.2cm
\\Let us denote by $\GG$ the class of locally finite
infinite connected graphs and by $\BB$ the class  of
bounded degree infinite connected graphs.
We further denote by $\QQ^v$ ($\QQ^e$) the class of vertex (edge) quasi-transitive graphs, and
by $\AA$ the class of amenable graphs.  In this paper we will not consider
non locally finite graphs.

\numsec=3\numfor=1
\section{The Model}

\\We define initially the model on a {\it finite}
graph $G=(V,E)$. For
each edge $e\in E$ we define a binary random variable $n(e)$,
which can assume the values $n(e)=1$ (open edge)
and $n(e)=0$ (closed edge). A configuration
$\om_G$ of the process is a function $\om: E\to \{0,1\}: e\mapsto n(e)$.
We call $\O_G$ the configuration space, i.e. the set of all
possible configurations of random variables $n(e)$ at the edges
$e\in E$ of the graph $G$. Given $\om\in \O_G$ we denote by
$O(\om)$ the subset of $E$ given by  $O(\om)=\{e\in E: \om(e)=1\}$
and by $C(\om)$ the
set $C(\om)=\{e\in E: \om(e)=0\}$. An {\it open connected component}
$g$ of $\om$ is
a connected subgraph $g=(V_g, E_g)$ of $G$ such that $E_g\neq\0$,
$\om(e)=1$ for all $e\in E_g$, and $\om(e)=0$ for all $e\in \partial g$.
A vertex $x\in V$ such that $\om(e)=0$ for all $e$ adjacent to
$x$ is an {\it isolated vertex} of $\om$.

The probability
$P_G(\om)$ to see the system in the configuration $\om\in \O_G$ is
defined as

$$
P_G(\om)=
{1\over Z_G(p,q)}
p^{|O(\om)|}(1-p)^{|C(\om)|} q^{k(\om)}
$$
where $p\in (0,1)$, $q\in(0,\infty)$, and $k(\om)$ is the number
of connected open components of the configuration $\om$ plus the number
of isolated vertices;
the normalization constant $Z_G(p,q)$, usually called the partition function
of the system,  is given by
$$
Z^{\rm RCM}_G(p,q)=
\sum\limits_{\om\in \O_G} p^{|O(\om)|}(1-p)^{|C(\om)|} q^{k(\om)}\Eq(ZG)
$$
The  ``pressure'' of the system is defined as the following function
$$
\pi_G(p,q)={1\over |V|}\ln Z^{\rm RCM}_G(p,q)
$$

\\In order to define the RCM on infinite graphs, we will
need to introduce the concept of boundary condition. Let
 $\GI=(\VU,\mathbb{E})$ a connected and locally finite infinite
graph and let $\O_\GI$ be the set of
all configurations in $\GI$, i.e. the set of all functions $\om$
such that $\om: \mathbb{E}\to \{0,1\}$.
Let $V\subset\VU$ a {\it finite} set and let $\GI|_V$ be
the restriction of $\GI$
to $V$. Given now $\x\in \O_\GI$, let $\O^\x_{\GI|_V}$ the
(finite) subset of $\O_\GI$
of all configurations $\om\in \O_\GI$ such that $\om(e)=\x(e)$
for $e\not\in \mathbb{E}|_{V}$.
For $\om \in \O^\x_{\GI|_V}$, let us also denote by $\omega_V$ the
restriction of $\omega$ on $\mathbb{E}|_{V}$. Note that $\om_V$ does
not depend on $\x$.
We now denote $P_{\GI|_V}^\x$ the random cluster probability measure  in $\O^\x_{\GI|_V}$
on the finite sub-graph $\GI|_V$ of the infinite graph $\GI$
with boundary conditions $\x$ as
$$
P_{\GI|_V}^\x(\om)={1\over Z^\x_{\GI|_V}(p,q)}
p^{|O(\om_V)|}(1-p)^{|C(\om_V)|}q^{k_V^\x(\om)}\Eq(Pmu)
$$
where $Z^\x_{\GI|_V}(p,q)$ is the partition function given by
$$
Z^\x_{\GI|_V}(p,q)=\sum_{\om\in \O^\x_V}p^{|O(\om_V)|}(1-p)^{|C(\om_V)|}
q^{k_V^\x(\om)}
\Eq(Zmu)
$$
and $k^\x_V(\om)$ is the number of {\it finite} connected open
component (open clusters) of the configuration
$\om$ (which agrees with $\x$ outside $V$) which intersect $V$
plus the number of isolated vertices in $V$.
Note that $k^\x_V(\om)$
is the only term in \equ(Pmu) and \equ(Zmu) depending
on boundary conditions
$\x$.

\\Two extremal boundary conditions play a central role, namely the
{\it free boundary} condition, in which $\x(e)=0$ for all $e\in \mathbb{E}$
and the {\it wired boundary condition},
in which $\x(e)=1$ for all $e\in \mathbb{E}$.
According to the definition above,
for a fixed configuration $\om$ with $\x=0$ outside
$V$ the number $k^0(\om)$  is actually the number of open components
in the finite sub graph $\GI|_V$ plus the isolated vertices in $V$,
while if $\x=1$ outside $V$,
all open components in $\GI|_V$ which touch the boundary
have not to be counted computing the number
$k^1(\om)$, since they belong to the infinite open cluster.
Thus $k^1(\om)$ is actually the number of finite
open connected component in $\om$ which do not touch the boundary
plus isolated vertices which do not belong to the boundary.

It is  important to remark here that in the above definition of $k^\x_V(\om)$
we compute only the {\it finite} connected components because we are
adopting the so called ``infinity-wired boundary condition'' convention,
see e.g. definition 2.1 in \cite{J} or section 2.3 in \cite{HJL1}. By this convention, all infinite open clusters eventually
intersecting $V$ are counted as one, i.e., as if all these clusters were connected
at infinity (wired at infinity).
In the literature one can also find  the so-called
``infinity-free boundary condition'' convention, in which
all open clusters, whether finite
or infinite, are counted in the number
$k(\om)$. In this case all infinite clusters intersecting $V$
are regarded as separate. This is e.g. the convention
adopted in the survey \cite{G2} and in the book \cite{G3}. In the rest of the paper
we will only consider the free ($\x=0$) and wired ($\x=1$ ) boundary conditions,
for which the ``infinity-free'' convention and the ``infinity-wired'' convention
are equivalent and we adopted the latter only because leads to
simpler definitions.

\begin{defi}\label{pressure}
Let $\GI=(\VU,\mathbb{E})\in \BB$;
let $\{V_N\}_{N\in\mathbb{N}}$ be a sequence of
finite subsets of  $\VU$ such that $V_N\nearrow \VU$ (not necessarily F\o lner);
let $\x$ be a boundary condition. Then we define, if it exists and it is
independent of $\{V_N\}_{N\in\mathbb{N}}$,  the
pressure of the random cluster model with parameters $q$ and $p$ and boundary condition
$\x$
on $\mathbb{G}$ as
$$
\pi^\x_{\mathbb{G}}(p,q)= \lim_{N\to \infty} {1\over  |V_N|} \ln Z^{\x}_{\GI|_{V_N}}(q)
\Eq(press)
$$
\end{defi}
In definition  \ref{pressure}, instead of choosing a fixed boundary
condition $\x$, one can also think to allow a whole sequence
$\x_N$  of boundary conditions, one for each $V_N\in \VU$.
However, as shown in \cite{Gr} (see also \cite{G2,G3}), this adds no extra generality.

\begin{rema}\label{mind}
With the further assumptions that $\GI$ is amenable, quasi-transitive and the sequence
$\{V_N\}_{N\in\mathbb{N}}$
is F\o lner, it is easy to prove that  this limit, which is known to exist for all $q>0$ and everywhere in the interval $p\in [0,1]$
except possibly in a countable set of points (see \cite{Gr,J}),
is independent of the boundary condition.
As a matter of fact, let $\x,\om\in \Omega_\GI$ and define $\om^{\x}_N$ by
$$
\om^{\x}_N(e)=\cases{ \om(e) &if $e\in \mathbb{E}|_{V_N}$\cr\cr
\x(e) & otherwise}
$$
Then, for all $\x$
$$
k^1_{V_N}(\om^{1}_N)\le k^\x_{V_N}(\om^{\x}_N)\le k^0_{V_N}(\om^{0}_N)\le
k^1_{V_N}(\om^{1}_N)+ |\partial V_N|
$$
whence
$$
Z^1_{{\GI|_{V_N}}}(p,q)\le Z^\x_{{\GI|_{V_N}}}(p,q)\le Z^0_{{\GI|_{V_N}}}(p,q)
\le
Z^1_{{\GI|_{V_N}}}(p,q)q^{|\partial V_N|}, ~~~{\rm if}~q\ge 1
$$
while for $q<1$ we have simply to reverse all inequalities above.
Now taking the logarithms, dividing by $|V_N|$, and using \equ(Folner)
one obtains the result.
\end{rema}

\\Other important quantities to study are the  so called
connectivity functions. To introduce them we need some preliminary
definitions.

\begin{defi}
\label{animal}Let $\GI\in \GG$.  {\it An animal in $\GI$ is a connected subgraph $g=(V_g, E_g)$ of $\GI$
with vertex set  $V_g$
and edge set $E_g$ such that $|V_g|<+\infty$ and $E_g\neq \0$.
We will denote by ${A}_\GI$
the set of all  animals in $\mathbb{G}$}.
\end{defi}

\begin{defi}\label{compa}
We say that two animals $ g_1=(V_{g_1},E_{g_1})$ and $ g_2=(V_{g_2},E_{g_2})$
in $\GI$ are compatible and we write $g_1\sim g_2$  if $V_{g_1}\cap V_{g_2}=
\emptyset$ (hence consequently
$E_{g_1}\cap E_{g_2}=\emptyset$). Otherwise we say that $g_1$ and $g_2$ are
incompatible and write
$g_1\not\sim g_2$.
\end{defi}

\\We are now ready to give the definition of connectivity functions.

\begin{defi}\label{connec}
Let $\GI=(\VU,\mathbb{E})\in \BB$ and let $X\subset \VU$ finite.
Let $\{V_N\}_{N\in\mathbb{N}}$ be a sequence of
finite subsets of  $\VU$ such that $V_N\nearrow \VU$ and  $X\subset V_N$
for all
$N\in \mathbb{N}$. Let $\x$ be a boundary condition. Then we define,
if it exists and it is
independent of $\{V_N\}_{N\in\mathbb{N}}$,  the
connectivity function  of the set $X$ of the random cluster model with parameters $q$ and
$p$ and boundary condition
$\x$
on $\mathbb{G}$ as
$$
\phi_{p,q,\x}(X)= \lim_{N\to \infty}
\sum_{\om\in \O^\x_{\GI_{N}}:~\exists g\in { A}_{\GI}:\atop E_g\in O(\om),~~
X\subseteq V_g} P_{\GI|_{V_N}}^\x(\om)
\Eq(conne)
$$

\\The
finite connectivity function of the set $X$ of the random cluster model
with parameters $q$ and $p$ and boundary condition $\x$
is defined as

$$
\phi^{\rm f}_{p,q,\x}(X)= \lim_{N\to\infty}
\sum_{\om\in \O^\x_{\GI_{N}}:~\exists g\in {A}_{\GI}:~ E_g\in O(\om)\atop
X\subseteq V_g,~V_g\cap ~\partial^{\rm int}_v V_N=\emptyset} P_{\GI|_{V_N}}^\x(\om)
\Eq(connef)
$$
\end{defi}
In the r.h.s of \equ(conne) the sum runs over configurations $\om$ containing an  animal made by open edges
whose vertex set contains $X$, while in the r.h.s of \equ(connef) the sum runs over configurations $\om$
containing an  animal made by open edges
whose vertex set contains $X$ and does not intersect the  boundary of $V_N$.

Let us define the subcritical phase of a RCM on a graph $\GI=(\VU,\mathbb{E})\in \BB$ at fixed $q$  as the set of values
of $p$ in the interval $[0,1]$ for which the probability  to find an infinite
open cluster in the system is zero. Conversely, the supercritical phase is the set of values
of $p$ in the interval $[0,1]$ for which the probability  to find an infinite
open cluster in the system containing a fixed vertex is strictly greater than zero.
\\We remark  that $\phi^{\rm f}_{p,q,\x}(X)$ coincides
with $\phi_{p,q,\x}(X)$  in the subcritical phase.

\\The connectivity function $\phi_{p,q,\x}(X)$
is expected to decay exponentially to zero when $d_\GI^{\rm tree}(X)\to\infty$ in the
subcritical phase, while, of course, is not expected to decay to zero in the supercritical phase,
where there is a non zero probability to find any set of vertices in the
infinite cluster. The exponential decay of the connectivity function in the
subcritical phase can  be obtained
for the RCM  on $\Z^d$ in the regime $q\ge 1$
by comparison inequalities  (see e.g. theorem 3.2 in \cite{G2})
and using the known results on Bernoulli bond percolation and/or Potts model.
On the other hand, the finite connectivity function
$\phi^{\rm f}_{p,q,\x}(X)$ is expected to decay exponentially to zero when $d_\GI^{\rm tree}(X)\to\infty$ in the
supercritical phase. Concerning again the RCM  on $\Z^d$ in the regime $q\ge 1$,
the exponential decay of finite connectivities (up to the slab percolation threshold in $d\ge 3$)
follows from the renormalization group analysis developed in \cite{Pi}.

\\It is well known (see e.g.  theorem 3.6 in \cite{G2}) that, for $q\ge 1$ we have, by FKG inequalities, that
$$\phi_{p,q,0}(X)\le\phi_{p,q,\x}(X)\le\phi_{p,q,1}(X)\Eq(dise1)$$
$$\phi^{\rm f}_{p,q,0}(X)\le\phi^{\rm f}_{p,q,\x}(X)\le
\phi^{\rm f}_{p,q,1}(X)\Eq(dise2)$$
for any boundary
condition $\x$.
Hence if one is able to prove  that
$$
\phi_{p,q,1}(X)=\phi_{p,q,0}(X),
$$
and/or
$$
\phi^{\rm f}_{p,q,1}(X)=\phi^{\rm f}_{p,q,0}(X),
$$
then automatically $\phi_{p,q,1}(X)=\phi_{p,q,\x}(X)=\phi_{p,q,0}(X)$
and/or  $\phi^{\rm f}_{p,q,1}(X)=\phi^{\rm f}_{p,q,\x}(X)=\phi^{\rm f}_{p,q,0}(X)$
for any fixed the boundary condition $\x$, {\it as far as $q\ge 1$}.
We stress that when  $q<1$ we cannot get to the same conclusion, since \equ(dise1) and
\equ(dise2) are false when $q<1$.

As it will be shown below we are able to prove using cluster expansion
techniques for all $q>0$ that
$\phi_{p,q,1}(X)=\phi_{p,q,0}(X)$  for $p$ sufficiently small and that
$\phi^{\rm f}_{p,q,1}(X)=
\phi^{\rm f}_{p,q,0}(X)$ for $p$ sufficiently near 1.
It is unclear for us if it is possible to  generalize
our expansions in order to include all boundary conditions in the whole regime $q>0$.
For these reasons we
preferred to treat only the simplest and most popular case $\x=0,1$.

\\Note finally that,  given a vertex $x_0\in \VU$,  the percolation
probability $\theta^\x_{p,q}(x_0\leftrightarrow\infty)$, i.e.
the probability that
there is an infinite open cluster passing through $x_0$ is defined in term
of the 1-point finite connectivity function as
$$
\theta^\x_{p,q}(x_0\leftrightarrow\infty)=
1- \phi^{\rm f}_{p,q,\x}(x_0)\Eq(theta)
$$

\\The critical percolation probability $p^\x_c(q)$ at a fixed value of
$q$ for the graph
$\GI$ is the value of $p$ defined by
$$
p^\x_c(q)= \sup_{p\in [0,1]\atop x_0\in \VU}
\{p: \theta^\x_{p,q}(x_0\leftrightarrow\infty)=0\}\Eq(pc)
$$
We recall that for the RCM on $\Z^d$ and $q\ge 1$
theorem 4.2  of \cite{A} states  that $p^\x_c(q)$ is independent of boundary conditions
and strictly smaller than $1$, while results of \cite{PS8} imply for
the RCM on $\Z^d$ with $q< 1$ that $p^{0,1}_c(q)<1$. We also recall that
for the particular case of $\Z^2$, duality arguments lead to the conjecture
that $p_c(q)=\sqrt{q}/(1+\sqrt{q})$. This conjecture has proven to be true for $q=1$ \cite{Ke},
$q=2$ \cite{O} and  for $q$ sufficiently large \cite{LMMRS}.

\vskip.5cm
\numsec=4\numfor=1

\numsec=4\numfor=1
\section{The subcritical phase}

\subsection{ Results in the subcritical phase}

\\We begin this section stating our two main
theorems about subcritical phase.
The first theorem concerns the connectivity functions.
The second concerns the pressure.
The rest of the section
will be devoted to the proof of these two theorems.

\begin{theo}\label{sub}
Let $\GI\in \BB$ with maximum degree $\D$. For any $q>0$, let $\{V_N\}_{N\in \mathbb{N}}$
be any sequence in $\VU$ such that
$V_N\nearrow \VU$ (it does not need to be F\o lner),
and let $p$ be so small that $(3+2\sqrt{2})\e_p\le 1$
where $$
\e_p=\max\left\{{e \D\over q}\left|{\ln (1-p)\over (1-p)^\D}\right|
, \,e \D\left|{\ln (1-p)\over (1-p)^\D}\right|\right\}\Eq(epsit)
$$

\\Then the infinite volume connectivity functions $\phi_{p,q,\x}(X)$
with  $\x=0,1$
of the
$RCM$ on $\GI$ defined in the limit \equ(conne) exist, are both equal
to a function $\phi_{p,q}(X)$ which can
be written explicitly in terms of an absolutely convergent
series which is analytic as a function of $p$, and does not
depend on the sequence $V_N$.

\\Moreover $|\phi_{p,q}(X)|$ admits the upper bound
$$
|\phi_{p,q}(X)|\le
{(7+5\sqrt{2})\over (2\sqrt{2}+3)}\left[\left(1+{1\over\sqrt{2}}\right)\, \e_p\,\right]^{d_\GI^{\rm tree}(X)-1}\Eq(exp1)
$$
where ${d_\GI^{\rm tree}(X)}$ is the tree distance of $X$ in $\GI$
accordingly to definition \ref{treed}.

\end{theo}

\begin{theo}\label{presub}
Let $\GI\in \BB\cap\AA\cap\QQ^v$ with maximum degree $\D$.
Let $q>0$ be fixed, let $\{V_N\}_{N\in \mathbb{N}}$
any F\o lner sequence in $\VU$ such that
$V_N\nearrow \VU$,
and let $p$ so small that $2 e^2\e_p^*< 1$
where $$
\e_p^*={e \D\over q}\left|{\ln (1-p)\over (1-p)^\D}\right|\Eq(epsits)
$$

\\Then
the pressure of Random Cluster Model on $\GI$, defined in \equ(press) exists
and can be written explicitly in term of an absolutely convergent
series which is analytic as a function of $p$, and  does not depend on $V_N$ and on $\x$.

\end{theo}

\\Note that the first theorem, concerning connectivity functions,
holds for a larger class of graphs, but
in a smaller region of parameters, while theorem \ref{presub} concerning
the pressure is valid for a smaller class of graphs,
which however includes all regular lattices, but in a larger region
of the parameters $p$ and $q$.

\\Once again we recall that the existence of these limits and independency of boundary
conditions is well known for the RCM on $\Z^d$ for $q\ge 1$ in the whole interval $p\in [0,1]$,
except in a subset at most countably infinite (conjectured to be a singleton or empty), see e.g. theorem 3.6 in \cite{G2}.

\subsection{Proof of theorem \ref{sub}. Polymer expansion
for the connectivity functions}

\\In this section we will assume that $\GI\in \BB$.
Let us take
sequence $\{V_N\}_{N\in \mathbb{N}}$  in
$\mathbb{V}$ tending monotonically to $\mathbb{V}$.
We will use the shorter notations
$\GI_N=\GI|_{V_N}$ and $\EE_N=\EE|_{V_N}$,
$k^\x_{V_N}=k^\x_N$
and also
$\omega_{\mathbb{E}_N}=\om_N$.

\\Fix a $X\subset V_N-  \partial_v^{\rm int}V_N$
(i.e., $X$ does not touch the boundary).
The finite volume free and wired connectivity function can be rewritten as
$$
\phi^N_{p,q, \x=0,1}(X)= {1\over \tilde Z^\x_{\GI|_N}(p,q)}
\sum_{\om\in \O^\x_{\GI_{N}}:~\exists g\in
{A}_{\GI}:\atop E_g\subset O(\om),~~
X\subset V_g}\l^{|O(\omega_N)|}
q^{k_N^\x(\om)}\Eq(Phi01)
$$

\\where
$$
\tilde Z^\x_{{\GI_{N}}}(p,q)=\sum_{\omega\in \Omega^\x_{\GI_N}}\l^{|O(\omega_N)|}
q^{k_{N}^\x(\om)}=(1-p)^{|\EE_N|}Z^\x_{{\GI_{N}}}(p,q)\Eq(Xi)
$$
and
$$
\l={p\over 1-p}\Eq(lambda)
$$

\\We recall that
$k_N^0(\om)$  is the number of open components
of $\om_N$ plus isolated vertices,
while $k_N^1(\om)$ is  the number of open connected component in $\om_N$
which do not intersect the boundary
plus isolated vertices which does belong to the boundary
$\partial_v^{\rm int}V_N$.

A configuration ${\omega\in \Omega^\x_{\GI_N}}$ is completely specified by
the set of open edges $O(\om_N)$ in $\EE_N$.
Let now $\{E_1, \dots , E_n\}$ be the connected components
of $ O(\om_N)$. To each $E_i$ we
can associate an animal $g_i\in \AA_{\GI_N}$
such that $V_{g_i}= \VU|_{E_i}$, $E_{g_i}= E_i$. Then to
each ${\omega\in \Omega^\x_{\GI_N}}$ can be associated a
(unordered) set of animals
$\{g_1, \dots , g_n\}_{\om_N}\subset\AA_{\GI_N}$
such that $\cup _{i=1}^n E_{g_i}= O(\om_N)$ and for all $i,j\in \setn$ ,
$g_i\sim g_j$. Observe that this one to one correspondence
$\om_N\leftrightarrow\{g_1, \dots, g_n\}$ yields
$$
|O(\om_N)|= \sum_{i=1}^n |E_{g_i}|\Eq(omn1)
$$
$$
\sum_{\omega\in \Omega^\x_{\GI_N}}(\cdot)=
\sum_{n\ge 0}\sum_{\{g_1,\dots ,g_n\}
\subset \AA_{\GI_N}\atop g_i\sim g_j}(\cdot)\Eq(somn)
$$
$$
\sum_{\om\in \O^\x_{\GI_{N}}:~\exists
g\in {A}_{\GI}:\atop E_g\subset O(\om),~~
X\subset V_g} (\cdot)=
\sum_{n\ge 1}\sum_{\{g_1,\dots ,g_n\}
\subset \AA_{\GI_N}\atop g_i\sim g_j, ~X\subset V_{g_1} }(\cdot)\Eq(somn2)
$$
where for $n=0$ the unordered $n$-uple  $\{g_1,\dots ,g_n\}$ is the empty set.

\\We will now rewrite the  partition function  \equ(Xi)
and the connectivity function \equ(Phi01)
in terms of the animals introduced above. We start by
considering the case $\x=0$.
Let us denote by $V_{\om_N}^{\rm iso}$ the
subset of $V_N$ formed by the isolated vertices in the configuration $\om_N$,
and let
$\{g_1, \dots , g_n\}_{\om_N}$
be the animals uniquely associated to $ O(\om_N)$. Then, by definition,
$$
k_N^0(\om)=n  + |V_{\om_N}^{\rm iso}|
$$
and since
$$
|V_{\om_N}^{\rm iso}|=
|V_N|-
 \sum_{i=1}^n|V_{g_i}|
$$
we obtain
$$
k_N^0(\om)= |V_N| - \sum_{i=1}^n\Big[|V_{g_i}|-1)\Big]
\Eq(k0)
$$
\\Using now \equ(omn1), \equ(somn), \equ(somn2) and \equ(k0),
the partition function
$\tilde Z^0_{{\GI_{N}}}(p,q)$  defined in \equ(Xi) can be rewritten as

$$
\tilde Z^0_{{\GI_{N}}}(p,q)=q^{|V_N|} \Xi^0_{{\GI_{N}}}(p,q)\Eq(xi0)
$$

\\where

$$
\Xi^0_{{\GI_{N}}}(p,q)= 1+
\sum_{n\ge 1}\sum_{\{g_1,\dots ,g_n\}\subset \AA_{\GI_N}\atop g_i\sim g_j}
\prod_{i=1 }^n {1\over q^{|V_{g_i}|-1}}\l^{|E_{g_i}|}
\Eq(Xii)
$$

\\and

$$
\phi^N_{p,q, \x=0}(X)= {1\over \Xi^0_{\GI_N}(p,q)}
\sum_{n\ge 1}\sum_{\{g_1,\dots ,g_n\}
\subset \AA_{\GI_N}\atop g_i\sim g_j, ~X\subset V_{g_1} }
\prod_{i=1 }^n {1\over q^{|V_{g_i}|-1}}\l^{|E_{g_i}|}
$$

\\The case $\x=1$ is slightly more involved. We first find an expression of
 $k_N^1(\om)$ in terms of the animals $\{g_1,\dots g_n\}$.
 The set $I_n=\{1,2,\dots, n\}$
is naturally partitioned in the disjoint union
of two sets $\setn^{\rm int}$ and $\setn^{\partial}$
defined as
$$
\setn^{\rm int}=\{i\in \setn: V_{g_i}\cap \partial_v^{\rm int}V_N=\0 \}
$$

$$
\setn^{\partial}= \{i\in \setn: V_{g_i}\cap \partial_v^{\rm int}V_N\neq\0 \}
$$

\\With these notations, denoting shortly
$V_N-\partial_v^{\rm int}V_N=V^{\rm int}_N$ and,
for $i\in \setn^{\partial} $,
$V_{g_i}^{\rm int}= V_{g_i}-\partial_v^{\rm int}V_N$, we have


%

$$
k_N^1(\om)= |V^{\rm int}_N| - \sum_{i\in  \setn^{\rm int} }(|V_{g_i}|-1)
-
\sum_{i\in \setn^{\partial} }|V_{g_i}^{\rm int}|
\Eq(k1)
$$

\\Hence in the case $\x=1$ we get

$$
\tilde Z^1_{{\GI_{N}}}(p,q)=q^{|V^{\rm int}_N|} \Xi^1_{{\GI_{N}}}(p,q)
$$
where
$$
\Xi^1_{{\GI_{N}}}(p,q)=
1+
\sum_{n\ge 1}\sum_{\{g_1,\dots ,g_n\}\subset \AA_{\GI_N}\atop g_i\sim g_j}
 \prod_{i\in \setn^{\rm int}} {1\over q^{|V_{g_i}|-1}}\l^{|E_{g_i}|}~
\prod_{i\in \setn^{\partial}}
{1\over q^{|V_{g_i}^{\rm int}|}}\l^{|E_{g_i}|}
$$

\\and

$$
\phi^N_{p,q, \x=1}(X)= {1\over \Xi^1_{\GI_N}(p,q)}
\sum_{n\ge 1}\sum_{\{g_1,\dots ,g_n\}\subset \AA_{\GI_N}
\atop g_i\sim g_j, ~X\subset V_{g_1} }
\prod_{i\in \setn^{\rm int}} {1\over q^{|V_{g_i}|-1}}
\l^{|E_{g_i}|}~ \prod_{i\in \setn^{\partial}}
{1\over q^{|V_{g_i}^{\rm int}|}}\l^{|E_{g_i}|}
$$
We now  rewrite $\phi^N_{p,q, \x=1}(X)$ in term of a polymer
expansion in which polymers are finite subsets of $\VU$ with cardinality greater than 1
which are said to be incompatible in the usual polymer expansion terminology if they overlap.

\\Let us now define, for each pair $\{x,y\}\subset \VU$,
$$
V_{xy}= \cases{ 0 & if $\{x,y\}\notin \EE$\cr\cr
\ln(1+\l) & if $\{x,y\}\in \EE$}
$$
Let us also define, for any subset $R\subset \VU$ such that $2\le |R|<+\infty$, the activity

$$
\r(R)=q^{-(|R|-1)}\sum\limits_{E'\subset {\rm P}_2(R)
\atop (R, E')\in {\cal G}_R}
\prod\limits_{\{x,y\}\in E'}(e^{V_{xy}}-1)\Eq(xir)
$$

\\where ${\cal G}_R$ is the set of connected graphs with vertex
set $R$.
For $R\subset V_N$ we also define  a $\x$-dependent  set activity as

$$
\r^\x(R)=\cases{ \r(R)
& if $\x=0$ \cr\cr
\r(R) & if  $\x=1$ and
$R\cap \partial_v^{\rm int} V_N=\0$\cr\cr
q^{-|R\cap V_N^{\rm int}|}\sum\limits_{E'\subset {\rm P}_2(R)\atop (R, E')
\in {\cal G}_R}
\prod\limits_{\{x,y\}\in E'}(e^{V_{xy}}-1) & if $\x=1$ and
$R\cap \partial_v^{\rm int} V_N\neq\0$
 }
\Eq(xirm)$$

\\Note that
$\r^0(R)$ is the restriction of $\r(R)$ for $R\subset \EE_N$ and
when $q<1$ we have, for all $R\in P_{\ge 2}(V_N)$, that

$$
|\r^\x(R)|\le |\r(R)| ~~~~~~~~~{\rm whenever} ~~q<1\Eq(indmu)
$$
Note also that
$$
\r^\x(R)=0~~~~~~~~~~ \mbox{whenever $R$ is not connected in $\GI$}
$$

We are thus ready to
define our polymer space.

\begin{defi}\label{subpoly}
We define the set of (subcritical) polymers as the set
$$
\PP=\{R\subset \VU: ~2\le |R|<+\infty,~~\mbox{$R$ is connected in $\GI$}\}
$$
We will say that  two polymers $R_i,R_j\in  \PP$  are
{\it compatible}, and we
write $R_i\sim R_j$, if
$R_i\cap R_j= \0$; viceversa,
$R_i$ and $R_j$  are
{\it incompatible}, and we
write $R_i\not\sim R_j$, if
$R_i\cap R_j\neq\0$. For $V_N\subset \VU$ finite we define
$$
\PP_N=\{R\subset \VU_N: |R|\ge 2,~~\mbox{$R$ is connected in $\GI$}\}
$$
\end{defi}

\\Then for $\x=0,1$ we can write

$$
\phi^N_{p,q, \x}(X)= {1\over \Xi^\x_{\GI|_N}(p,q)}
\sum_{n\ge 1}{1\over n!}\sum_{(R_1,\dots , R_n)\in \PP_N^n\atop
R_i\sim R_j,~\exists i\in{\rm I}_n:\;
R_i \supset X} \r^\x(R_1)\cdots\r^\x(R_n)\Eq(phim)
$$
where ${\rm I}_n=\{1, 2, \cdots, n\}$ and $\PP^n$
is the $n$-times cartesian product of
$\PP$, i.e. elements
of $\PP_N^n$  are ordered $n$-ples of elements of $\PP_N$.
The partition function $\Xi^\x_{\GI|_N}(p,q)$ can be rewritten as

$$
\Xi^\x_{{\GI_{N}}}(p,q) =
\Bigg[1  +\sum_{n\ge 1}{1\over n!}\sum_{(R_1,\dots , R_n)\in \PP_N^n\atop
R_i\sim R_j} \r^\x(R_1)\cdots\r^\x(R_n)\Bigg]
\Eq(XIm)
$$
The factor 1 in r.h.s.
is the contribution of the configuration in which all edges in $\GI_N$ are closed.
Observe that  the partition function is rewritten as a genuine Gruber and Kunz
hard core polymer gas partition function in which the polymers
are finite  subsets $R$ of $V_N$ with cardinality greater than one and with activity
$\r^\x(R)$.

\\It is now easy to rewrite this ratio (between two finite sums) as an infinite series.
Define, for $R\in \PP$

$$
\Pi^N_{p,q, \x}(R)={\partial\over \partial \r^\x(R)}\ln\Big[ \Xi^\x_{{\GI_{N}}}(p,q)  \Big]
$$

\\Then, by construction
\def\gt{{\tilde \g}}
\def\Gt{{\tilde\Gamma}}
$$
\phi^N_{p,q, \x}(X)=\sum_{R\in \PP_N\atop
X\subset R}\r^\x(R)\Pi^N_{p,q, \x}(R)\Eq(PhiPiN)
$$
Now, by standard cluster expansion it is  well known that
$$\ln \Xi^\x_{{\GI_{N}}}(p,q)  =
\sum_{n\ge 1}{1\over n!}\sum_{(R_1,\dots , R_n)\in \PP_N^n}
\r^\x(R_1)\cdots\r^\x(R_n)
\F^T (R_1,\dots , R_n)
\Eq(log)
$$

\\where the Ursell coefficients
$\F^T (R_1,\dots , R_n)$ are given by
$$
\Phi^{T}(R_1,\dots , R_n)=
\cases{\sum\limits_{E\subset E(R_1,\dots,R_n)\atop(\setn,E)\in \GG_n}(-1)^{|E|}
&if $n\ge$ 2\cr\cr\cr
1&if $n=1$.}
\Eq(PhiTU)
$$
where $E(R_1,\dots,R_n)=\{\{i,j\}\subset \setn: R_i\not\sim R_j\}$ and
$\GG_n$ denotes the set of all connected graphs with vertex set $\setn$.
So
$$
\Pi^N_{p,q, \x}(R)=\sum_{n\ge 0}{1\over n!}\sum_{(R_1,\dots , R_n)\in \PP_N^n}
\r^\x(R_1)\cdots\r^\x(R_n)
\F^T (R,R_1,\dots , R_n)\Eq(PiN)
$$

\\We also define
functions  on the whole $\GI$
(hence not depending on boundary conditions) as follows
$$
\Pi_{p,q}(R)=\sum_{n\ge 0}{1\over n!}\sum_{(R_1,\dots , R_n)\in \PP^n}
\r(R_1)\cdots\r(R_n)
\F^T (R,R_1,\dots , R_n)\Eq(Pi)
$$

$$
\phi_{p,q}(X)=
\sum_{R\in \PP\atop
X\subset R}\r(R)\Pi_{p,q,}(R)
\Eq(PhiPi)
$$
We can now use the methods of the abstract polymer gas, see \cite{KP,FP} to determine the
convergence radius  for the series \equ(PiN) and \equ(Pi) and their bounds.
We will see that this formal series are indeed an absolutely
convergent expansions for the infinite volume connectivity functions
for $p$ sufficiently small.

\subsection{Proof of theorem \ref{sub}.
Convergence of the connectivity functions}

\\First we prove an exponential bound on the activity $\r(R)$, which is an essential ingredient for the convergence of the cluster
expansion.

\begin{lem}\label{actysub}
Let $\GI\in \BB$ with maximum degree $\D$. Then, for any $n\ge 2$
and $\x=0,1$
$$
\sup_{x\in \mathbb{V}}\sum_{R\in \PP:~ \atop x\in R,~ |R|=n}|\r(R)|
\le (\e_p^*)^{n-1}\le \e_p^{n-1} \Eq(bfk1)
$$
and,
$$
\sup_{x\in V_N}\sum_{R\in \PP_N\atop x\in R,~ |R|=n}|\r^\x(R)|
\le \e_p^{n-1} \Eq(bfk2)
$$
where
$\e_p$ and $\e_p^*$ are defined in \equ(epsit) and \equ(epsits) respectively.
\end{lem}

\\{\bf Proof.} Observe that, for $R\in P_{\ge 2}(\VU)$ by
definition of \equ(xir)
$$
\sup_{x\in \mathbb{V}}\sum_{R\in P_{\ge 2}(\VU):~ x\in R\atop |R|=n}|\r(R)|\le
|q|^{-(n-1)}\sup_{x\in \mathbb{V}}
\sum_{R\in P_n(\VU)\atop x\in R}\Bigg|
\sum\limits_{E'\subset {\rm P}_2(R)\atop (R, E')\in {\cal G}_R}
\prod\limits_{\{x,y\}\in E'}  \Big[e^{V_{xy} }-1\Big]\Bigg|\Eq(fkk)
$$
while, for $\r^\x(R)$ we have in the worst case (i.e. for
$R\subset \partial_v V_N^{\rm int}$)
$$
\sup_{x\in V_N}\sum_{R\in P_{\ge 2}(V_N)\atop x\in R,\; |R|=n}|\r^\x(R)|\le
\sup_{x\in \mathbb{V}}
\sum_{R\in P_{n}(\VU):\atop x\in R}\left|
\sum\limits_{E'\subset {\rm P}_2(R)\atop (R, E')\in {\cal G}_R}
\prod\limits_{\{x,y\}\in E'}  [e^{V_{xy} }-1]\right|\Eq(fkkmu)
$$
Then all we have to show to prove the lemma is that
$$
\sup_{x\in \mathbb{V}}
\sum_{R\in P_{n}(\VU):\atop x\in R}\left|
\sum\limits_{E'\subset {\rm P}_2(R)\atop (R, E')\in {\cal G}_R}
\prod\limits_{\{x,y\}\in E'}  [e^{V_{xy} }-1]\right|
\le (e |f_\D(p)|)^{n-1}
$$

\\Using thus the Battle-Brydges-Federbush inequality (see
e.g. \cite{Br}), recalling that
$\mathbb{E}|_R=\{\{x,y\}\in \mathbb{E}: x\in R, y\in R\}$, and observing that
$\sum_{\{x,y\}\in R} V_{xy}\le {1\over 2} \D |R|\le \D(|R|-1)$
for all $R$ such that $|R|\ge 2$, we get
$$
\Bigg|\sum\limits_{E'\subset {\rm P}_2(R)\atop (R, E')\in {\cal G}_R}
\prod\limits_{\{x,y\}\in E'}  [e^{V_{xy} }-1]\Bigg|
\le
[(1+\l)^{\D} \ln(1+\l)]^{|R|-1}\sum\limits_{E'\subset {\rm P}_2(R)
\atop (R, E')\in {\cal T}_R}
\prod\limits_{\{x,y\}\in E'} \d_{|x-y|1}
$$
where ${\cal T}_R$ is the set of all connected {\it tree graphs} with
vertex set $R$
and $\d_{|x-y|1}=1$ if $|x-y|=1$ and $\d_{|x-y|1}=0$ otherwise. It is now easy to check that

$$
\sum\limits_{E'\subset {\rm P}_2(R)
\atop (R, E')\in {\cal T}_R}
\prod\limits_{\{x,y\}\in E'} \d_{|x-y|1}
\le
\sup_{x\in \mathbb{V}}
\sum_{R\in P_{n}(\VU):\atop x\in R}
\sum\limits_{E'\subset {\rm P}_2(R)\atop (R, E')\in {\cal T}_R}
\prod\limits_{\{x,y\}\in E'}  \d_{|x-y|1}\le
$$

$$
\le {1\over (n-1)!}
\sum\limits_{E'\subset {\rm P}_2(\setn)\atop (\setn, E')\in {\cal T}_n}
\Bigg[\sup_{x\in \mathbb{V}}\sum_{x_1=x,\,(x_2,\dots , x_n)\in \VU^{n-1}
\atop x_i\neq x_j~\forall \{i,j\}\in \setn}
\prod\limits_{\{i,j\}\in E'}  \d_{|x_i-x_j|1}\Bigg]
$$
Now observe that, for any $E'\subset {\rm P}_2(\setn)$ such that
$(I_n, E')$ is a tree,
it holds
$$
\sup_{x\in \mathbb{V}}\sum_{x_1=x,\,(x_2,\dots , x_n)\in \VU^{n-1}
\atop x_i\neq x_j~\forall \{i,j\}\in \setn}
\prod\limits_{\{i,j\}\in E'}  \d_{|x_i-x_j|1}\le {\D^{n-1}}
$$
Moreover,using Cayley formula,
$|\{E'\subset {\rm P}_2(\setn):~ (R, E')\in {\cal T}_n\}|=
n^{n-2}$,  and the estimate ${n^{n-2}/ (n-1)!}\le e^{n-1}$, we can conclude that

$$
\sup_{x\in \mathbb{V}}
\sum_{R\in P_{n}(\VU):\atop x\in R}\Bigg|
\sum\limits_{E'\subset {\rm P}_2(R)\atop (R, E')\in {\cal G}_R}
\prod\limits_{\{x,y\}\in E'}  [e^{V_{xy} }-1]\Bigg|
~\le~
\left[e\D(1+\l)^{\D} \ln(1+\l)]\right]^{(n-1)}
$$
$\Box$

\vv
\\Using this result one can the prove the following lemma

\begin{lem}\label{phiNan}
For any $q>0$, the function $\phi_{p,q}(X)$ defined in \equ(PhiPi) is
analytic as a function of $p$  whenever
 $(3+2\sqrt{2})\e_p\le {1}$ where $\e_p$ is the number  in \equ(epsit) and satisfies
 the bound \equ(exp1), uniformly in $V_N$ and $\x=0,1$.
Moreover the function $\phi^N_{p,q, \x}(X)$ defined in \equ(Phi01) is also analytic as a function of $p$
whenever $(3+2\sqrt{2})\e\le {1}$ and $|\phi^N_{p,q, \x}(X)|$ is bounded above by the r.h.s. of \equ(exp1).
\end{lem}

\\{\bf Proof}. Using the condition (3.16) of \cite{FP}, valid  for polyemers
whose incompatibility relation is the overlapping, we have that the series \equ(Pi) converges if
$$
\sup_{x\in \VU}\sum_{R\in \PP\atop x\in R}|\r(R)|e^{a|R|}\le e^a-1\Eq(FP)
$$
Using lemma \ref{actysub} we have that
$$
\sup_{x\in \VU}\sum_{R\in \PP\atop x\in R}|\r(R)|e^{a|R|}~\le~ \sum_{n\ge 2} e^{a|n|}
\sup_{x\in \VU}\sum_{R\in \PP\atop x\in R: |R|=n}|\r(R)|~\le~  \sum_{n\ge 2} e^{a|n|}\e^{n-1}
$$
So condition \equ(FP) is optimal for $a=\ln(1+{1\over \sqrt{2}})$ and gives
$$
\e\le {1\over 3+2\sqrt{2}}\Eq(OKK)
$$
This for $\e$ satisfying \equ(OKK) the series \equ(PiN) and \equ(Pi) are convergent
and, by theorem 1 of \cite{FP} (see there formula (3.17)) we have the bound
$$
\Pi_{p,q,}(R)\le  e^{a|R|}\le \left(1+{1\over \sqrt{2}}\right)^{|R|}
$$
So, recalling \equ(PhiPiN) and \equ(PhiPi)
and observing that $\min\{|R|: R\in \PP,~X\subset R\}=d_\GI^{\rm tree}(X)$, we get
$$
|\phi_{p,q,}(X)|=\sum_{R\in \PP\atop
X\subset R}|\r(R)|\left(1+{1\over \sqrt{2}}\right)^{|R|}=
\sum_{n\ge d_\GI^{\rm tree}(X)}\e_p^{n-1} \left(1+{1\over \sqrt{2}}\right)^{n}\le
$$
$$
\left(1+{1\over \sqrt{2}}\right)\sum_{n\ge d_\GI^{\rm tree}(X)-1}
\left[\e_p\left(1+{1\over \sqrt{2}}\right)\right]^{n}\le
{(7+5\sqrt{2})\over (2\sqrt{2}+3)}\left[\e_p(\sqrt{2}+1)\over\sqrt{2} \right]^{d_\GI^{\rm tree}(X)-1}
$$
The proof that $\phi^N_{p,q, \x}(X)$ is also analytic and $|\phi^N_{p,q, \x}(X)|$  admits the same upper bound
\equ(exp1) is completely analogous just
observing that,
by \equ(bfk2) and \equ(epsit),
$\sup_{x\in \VU}\sum_{R\ni x: |R|=n}|\r^\x(R)|$ admits
the same bound of $\sup_{x\in \VU}\sum_{R\ni x: |R|=n}|\r(R)|$. $\Box$

\\Finally we prove the following result which ends the proof
of theorem \ref{sub}.
\vv

\begin{lem}\label{lim}
Let $\GI=(\VU, \EE)$ be a bounded degree graph
and let
$\{V_N\}$  be any sequence in $\VU$  such that
$V_N\nearrow\VU$ .
Then for any fixed $q>0$, $\x=0,1$ and $p$ such that $(3+2\sqrt{2})\e_p\le {1}$

$$
\lim_{N\to \infty} \phi^N_{p,q,\x}(X) = \phi_{p,q}(X)
$$
where $\phi_{p,q}(X)$ is the function defined in \equ(PhiPi).
\end{lem}

\\To prove this theorem we will first need to prove   a simple graph theory lemma stated as follows.

\begin{lem}\label{graph}
Let $\GI=(\VU,\EE)$ be bounded degree, let $V_N\nearrow \VU$ be a sequence of
finite subsets tending monotonically to $\VU$, and let $x$ a
vertex of $\GI$ such that
$x\in V_N$  for all $N$, then
$$
\lim_{N\to\infty} d(x, \partial_v^{\rm int} V_N)=+\infty
$$
\end{lem}

\\{\bf Proof}. Suppose that it is possible to find $x_0\in \bigcap_N V_N$
such that
$d(x_0, \partial_v^{\rm int} V_N)<R$ for some real constant $R$.
Then one can construct an
infinite sequence $\{x_N\}_{N\in \N}$ of {\it distinct} vertices
such that $x_N\in V_N$ but $x_N\notin V_{M}$ for all
$M<N$ and $d(x_0, x_N)\le R$ for all $x_N$. So this means that
all $x_N$ are in the ball of radius $R$
and center $x_0$. But since $\GI$ is bounded degree this ball is
finite and we have a contradiction. $\Box$

\\We are now ready to prove the lemma \ref{lim}.
\vv
\\{\bf Proof of lemma \ref{lim}}.
Let us consider the case $\x=1$,  which is the less trivial case.
$$
|\phi_{p,q}(X)- \phi^N_{p,q,\x=1}(X)|=
$$

$$
=~\sum_{n\ge 1}{1\over (n-1)!}\sum_{(R_1,\dots , R_n)\in \PP^n \atop
X\subset R_1, \exists j:~ R_j\not\subset V_N}
\r(R_1)\cdots\r(R_n)
\F^T (R_1,\dots , R_n)~+
$$

$$
+~\sum_{n\ge 1}{1\over (n-1)!}\sum_{(R_1,\dots , R_n)\in \PP_N^n \atop
X\subset R_1, \exists j:~ R_j\cap \partial_v^{\rm int}V_N\neq\0}
|\r(R_1)\cdots \r(R_n)-\r^1(R_1)\cdots \r^1(R_n)|
\F^T (R_1,\dots , R_n)
$$

Now, the first term of the r.h.s. of this inequality is,
for $(3+2\sqrt{2})\e_p\le {1}$,  clearly at least of the order
$([1+1/\sqrt{2}]\e_p)^{d_\GI(X, \partial_v^{\rm int} V_N)}$, with since
one among the $R_1, \dots, R_n$ has to contain $X$
and another has to intersect $\VU-V_N$. Recall that
the sets $R_1, \dots, R_n$ are pairwise
intersecting due to the presence of the factor $\F^T ( \ER_n )$.

\\The second term can be treated similarly, due to the
bounds \equ(bfk1) and \equ(bfk2),
and again one shows  that it is of the order
$([1+1/\sqrt{2}]\e_p)^{d_\GI(X, \partial_v^{\rm int} V_N)}$.
Now as $N\to \infty$ we have clearly that
$d_\GI(X, \partial_v^{\rm int} V_N)\to \infty$ due to lemma
\ref{graph}. The proof of the case $\x=0$ is the same, since
just the first term in the inequality above is present.
 $\Box$

\subsection{Proof of theorem \ref{presub}}
\\To prove theorem \ref{presub}, we recall that
the pressure of the random cluster model
is given by \equ(press). As it has been shown in the remark \ref{mind},
if the pressure exists, it is independent of boundary conditions. Hence we can work here with free boundary conditions
$\x=0$ which are easier for small $p$.

\\Now by \equ(Xi) and \equ(xi0)

$$
{1\over  |V_N|} \ln Z^{0}_{\GI|_{V_N}}(q)={1\over  |V_N|}
\ln \Xi^{0}_{\GI|_{V_N}}(q)
 -{|\EE_N|\over V_N}\ln (1-p) +  \ln q
$$

\\where we recall that  $\Xi^\x_{{\GI_{N}}}(p,q)$ is given
explicitly by equation \equ(XIm).

\\We have

\begin{prop}\label{orbit}
Let $\GI$ amenable and quasi-transitive with vertex orbits $O_1, \dots , O_k$,
let $\D_i$ be the degree of the vertices in the orbit $O_i$
(for $i=1,\dots, k$), and let $\{V_N\}_{N{\in}\mathbb N}$
be a F\o lner sequence such that $V_N\nearrow\VU$. Then,
there exists a non-zero finite limit
$$
\lim_{N\to \infty}{ |\EE _{N}|\over |V_N|}\Eq(limo)
$$
independent of the choice of the F\o lner
sequence $\{V_N\}_{N{\in}\mathbb N}$.
\end{prop}

\\{\bf Proof.} By  lemma 6 of \cite{PSG}  the limit
$$
\lim_{N\to\infty}{|O_i\cap V_N|\over |V_N|}= \a_i
$$
exists and it is independent of the choice of the sequence
$\{V_N\}_{N{\in}\mathbb N}$. Hence, considering that
each vertex in an orbit $O_i$ has $\D_i$ edges and each of
these edges counts 1/2 since
it is shared with another   vertex, one obtains immediately that
$$
\lim_{N\to \infty}{ |\EE _{N}|\over |V_N|}= {1\over 2}
(\a_1 \D_1 + \dots + \a_k\D_k)\Eq(EnVn)
$$

\\$\Box$

\vv

\\By this proposition we have that

$$
\pi_{\mathbb{G}}(p,q)=
\lim_{N\to \infty} {1\over  |V_N|} \ln \Xi^{\x}_{\GI_N}(q)
-{1\over 2}(\a_1 \D_1 + \dots + \a_k\D_k)\ln(1-p) + \ln q
$$

\\Thus in order to show that the pressure exists  we need to prove
that the limit
$$
\Pi_\GI(p,q)=\lim_{N\to \infty} {1\over  |V_N|} \ln \Xi^{0}_{\GI_N}(q)\Eq(gasp)
$$

\\exists, is independent of $V_N$ and
has a finite radius of convergence.

\\By the previous analysis, when the condition \equ(FP) is satisfied, the
logarithm of $\Xi^{0}_{\GI_N}(p,q)$ converges absolutely, and we can use as an estimate of its
radius of convergence $\e_p^*$ instead of $\e_p$, since we are using for the computation of the
pressure free boundary conditions. This ends the proof of theorem \ref{presub}.
$\Box$

\numsec=5\numfor=1
\section{The supercritical phase}

\subsection{More definitions about graphs and the main results in
the supercritical regime}

\\In order to study the supercritical phase we  need
to introduce the concept of cut-sets and minimal cut-sets of a
graph. We will define a  special  class
of minimal cut-sets in an infinite  graph which may be regarded as
the generalization of the
concept of Peierls contours used in the Potts
model defined in $\Z^d$.
We recall that a {\it cut-set} of a graph $\GI\in\cal G$ is
a set $\g\subset \mathbb{E}$ such that the graph $(\VU, \EE-\g)$ is disconnected.

\begin{defi}\label{fence}
A finite  cut-set $\g$ of an infinite connected graph
$\GI=(\VU,\EE)\in \GG$ is called a fence if
$(\VU, \EE-\g)$  has one and only one finite
connected component and for all edges $e\in \g$
the graph $(\VU, \EE-(\g-e))$  has no finite
connected component. If $\g$ is a fence, we denote by $g_\g=(I_\g , E_\g)$ the
unique finite connected component of $(\VU, \EE-\g)$.
The set $I_\g\subset \VU$ is called {\it the vertex interior of the fence $\g$}, and
$O_\g= \VU-I_\g$ is called {\it the vertex exterior of the fence $\g$}.
Analogously the set $E_\g\subset \EE$ is called {\it the edge interior of the fence $\g$}, and
$\EE_\g= \EE-\{\g\cup E_\g\}$ is called {\it the edge exterior of the fence $\g$}.
\end{defi}

\\Note that for any  fence $\g$ of $\GI=(\VU,\EE)$ it follows directly from the definition
that $I_\g\cap O_\g=\0$ and $I_\g\cup O_\g=\VU$. Moreover $\g\cap E_\g=\g\cap \EE_\g=E_\g\cap\EE_\g=\0$
and $E_\g\cup\g\cup\EE_\g=\EE$. From definition \ref{fence} it also follows that $\partial_e I_\g=\g$,  $E_\g=\EE|_{I_\g}$ and
$\EE_\g=\EE|_{O_\g}$. Moreover, any edge $e\in \g$ is such that $e=\{x,y\}$ with $x\in I_\g$ and $y\in O_\g$.
If $\g\subset \EE$ is a fence,
we put  $\GI_\g=(O_\g, \EE_\g)$. Note that $\GI_\g$ is an infinite graph but in general it is not connected.
We finally denote by  $\G_{\GI}$ the set of all fences in $\GI$.

\\A slightly less immediate property of fences is given by
the following proposition which shows that a fence $\g$ is, $\forall v\in I_\g$,
a $(v,\infty)$-minimal cut-set in the sense of \cite{BB}.

\begin{prop}\label{ray}
Let $\g$ be a fence in $\GI$  and let $x\in I_\g$, then
for any ray $\r=(V_\r, E_\r)$ in $\GI$ starting at $x$ we have that
$E_\r\cap \g\neq\emptyset$.
\end{prop}

\\{\bf Proof.} Suppose by contradiction that $E_\r\cap \g=\emptyset$. Then
$E_\r\subset E^1_\r\cup E^2_\r$
with $E^1_\r\subset E_\g$ and $E^2_\r\subset \tilde\EE_\g$
where
$ {\tilde \GI}_\g=(\tilde O_\g, \tilde\EE_\g)$
is some (infinite) connected component of
$\GI_\g$.
The case  $E^2_\r=\emptyset$ would imply that $E_\r\subset E_\g$ which is
impossible since $E_\r$ is infinite and $E_\g$ is finite. The case $E^1_\r=\emptyset$
is impossible since no edge in $\EE_\g$ has $x$ as one of its end-points.
Finally the last case $E^1_\r\neq\emptyset$ and $E^2_\r\neq\emptyset$
is impossible since otherwise $g_\g\cup\tilde\GI_\g\subset (\VU, \EE-\g)$ would be connected and infinite which
contradicts definition \ref{fence}. $\Box$

\vv

\\We will also use the following definitions:

\begin{defi}\label{surround}
Given a fence $\g \subset {\mathbb{E}}$ and a vertex set
$X\subset\VU$, we say that $\g$ surrounds $X$ and we write
$\g\bigodot X$ if
$X\subset I_\g$. We say that {\it $\g$ separates $X$}
and we write $\g\bigotimes X$,
if for any animal ${a}=(V_a,E_a)$ such that
$X\subset V_a$, then $E_a\cap \g\neq \emptyset$.
\end{defi}

\begin{defi}\label{restri}
Let $\GI=(\VU, \EE)\in \GG$, let $V\subset \VU$ and let $R\ge 1$.
We define the graph $\GI|_V^R$ as the graph with vertex set $V$ and edge set
$E=\{\{x,y\}:\,x,y\in  V\,\, {\rm and}\,\, d_\GI(x,y)\le R\}$.
$V\subset \VU$ is called $R$-connected if $\GI|_V^R$ is connected.
Analogously a set $S\subset \EE$ is
$R$-connected if its support $V_S$ is $R$-connected.
\end{defi}
\\In other words a set $V\subset \VU$ is $R$-connected in $\GI=(\VU, \EE)$, if for any partition
$\{A,B\}$ of $V$ such that $A\cap B=\0$ and $A\cup B=V$ we have that $d_\GI(A,B)\le R$.

\begin{defi}\label{boundcut}
A graph $\GI\in \GG$ is called cut-set-bounded if there exists
$R<+\infty$ such
that every fence $\g$ in $\GI$ is $R$-connected. We denote by $\PP$ the
subclass of $\GG$ of all
cut-set-bounded  graphs. Given a cut-set-bounded graph $\GI$ we call
the constant
$$
R_{\GI}=\min\{ R\in \mathbb{R}:\;\mbox{every cut-set is $R$ connected} \}
\Eq(cutsetc)
$$
the cut-set constant of $\GI$.
\end{defi}

\def\WW{{\mathcal W}}

\begin{defi}\label{non root}.
Let  $\GI$ be locally finite  graph, and let, for any $n\in \N$
$$
\WW_n=\{W\subset \VU: |W|<\infty\,,\mbox{ $W$ connected},\,{\rm diam}(W)=n \}
$$
We define the function $f_\GI: \N\to \N$ with
$$
f_\GI(n)=\min_{W\in \WW_n}|\partial W| \Eq(fGI)
$$
so that
$$
|\partial_e W|\ge f_\GI({\rm diam}(W) )  ~~~\mbox{for all $W\subset \VU$
finite and connected}\Eq(iib).
$$
The function $f_\GI$ is called  {\it the cut-set function of the graph}.
\end{defi}
Roughly speaking, this function measure how, in a graph $\GI$, the boundary of connected sets of minimal boundary
grows with the diameter of the set. Note that, by definition, $f_\GI$ grows at most linearly with $n$ in any bounded degree
graph. Indeed, for  most of the  known examples (e.g. $\Z^d$ and regular trees)  $f_{\GI}$ is a linear function.
To construct an example of $\GI$ for which $f_{\GI}$  grows slower than linearly, e.g. as $\ln n$, consider the infinite subset of $\Z^2$
below the curve $\ln x$ and above the $x$-axis. It is not difficult to see that such a graph has sets of diameter $n$ that can be
disconnected from the graph by deleting $\ln n$ edges.

\begin{defi}\label{non root2}
An infinite graph  $\GI$
is called a percolative graph if $\GI\in \PP\cap\BB$ and its cut-set function $f_\GI$ admits the lower bound
$$
f_\GI(n)\ge C\ln n\Eq(iib2)
$$
for some constant $C$.
We denote by $\LL$ the set of percolative graphs.
\end{defi}

\\We refer to
graphs satisfying definition above  as percolative because, as we will see below,
the conditions in definition \ref{non root2} are
sufficient  conditions for a graph
to exhibit a non trivial percolation threshold.
Heuristically, the requirement that the graph belongs to the class $ \PP\cap\BB$ is a sufficient
condition for the number of  fences
(i.e. the analogous of contours of the Ising model in $\Z^d$) of size $n$ containing a fixed edge
to grow at most as $C^ n$, while the condition \equ(iib2) is enough to guarantee that the number of possible
positions of fences of size $n$ surrounding a fixed vertex can be at most $C^n$ (which occurs when $f_\GI\sim \ln n$).
We remark that our conditions are far from being necessary. For example, the class of graphs $\PP\cap\BB$ does not contain the
trees (trees have fences which are not $R$-connected for any finite $R$) which do exhibit a non trivial percolation threshold.

\vv

\\To study the infinite  volume limit of the connectivity functions
in  percolative graphs and in particular
to ensure independence of this limit from boundary conditions  $\x=0,1$,
we will need to slightly restrict the class of sequence
$\{V_N\}$ along which this limit is taken. So we have to introduce one
more definition.

\begin{defi}\label{seque}
Let $\GI=(\VU, \EE)\in \PP$ with cut-set function $f_\GI(n)$ and let $\{V_N\}$ a
sequence of subsets of $\VU$ such that
$V_N\nearrow\VU$; we say that $V_N$ is a cut-set bounded sequence if for
all $N$ and for all
fences $\g$ such that $V_N\cap I_\g\neq\0$,  we have that the edge
set $\g\cap\EE_N$ is $R$-connected where $R$ is the cut-set constant of $\GI$.
\end{defi}
\\We were not able to find a graph $\GI=(\VU,\EE) \in \PP$ which does not admit a cut-set-bounded sequence of
sets $V_N$ invading $\VU$. Roughly speaking one should be able to produce example of graphs $\GI=(\VU,\EE)$ in $\PP$
with cut-set constant $R$ such that,  given any finite set $V\subset \VU$  there are fences $\g$ of $\GI$ such that
$\g\cap \EE|_V$ is not $R$-connected. On the  other hand,
we were also not able to prove that if $\GI\in \PP$ then it always exists  such a sequence.

\\We are now in the position to state our  results  concerning the supercritical
regime of the Random
Cluster model with free or wired boundary conditions and for $p$ sufficiently
close to 1. These results
will be resumed by stating two theorems, the first concerning the finite
connectivity functions and
the second concerning the pressure.
We remind that in the supercritical phase the interesting quantities are the
{\it finite} connectivity functions (see comments after definition \ref{connec} and, for $q=1$,
see also \cite{G1} )   defined in \equ(connef). That is why
the theorem \ref{sup} below will be stated in term of these
quantities.

\begin{theo}\label{sup}
Let $\GI=(\VU,\EE)\in \LL$
with cut-set constant $R$, let $\{V_N\}_{N\in \mathbb{N}}$ be
any cut-set bounded sequence in $\VU$ such that
$V_N\nearrow\VU$,
let $q>0$ be fixed,
and let $(1-p)$ so small   that  $eA(1+\D^{R+1})\d_p\le 1$ where
$\Delta$ is the maximum degree of $\GI$ and

$$
A=\Big[\max\{2C, 1\}\Big] \times
\Big[\D^{2R}\Big]\Eq(deltat0)
$$

$$
\d_p=  \max\Bigg\{ \Big|{1-p\over p}\Big| q, \Big|{1-p\over p}\Big| \Bigg\}\Eq(defel0)
$$

\\Then:

\\i) the infinite volume connectivity functions of the
$RCM$ on $\GI$ with free and wired boundary conditions,
defined in the limit \equ(conne), exist and
are both equal to a function $\phi^{\rm f}_{p,q}(X)$ which
can be
written explicitly in term of an absolutely convergent
series analytic as a function of $p$ near 1, and does not depend on
the sequence $V_N$.
\vskip.3cm
\\ii)
$|\phi^{\rm f}_{p,q}(X)|$ admits the bound

$$
|\phi^{\rm f}_{p,q}(X)|~\le~
(1+\D^{-R-1}) (Ae\,\d_p\,)^{f_\GI( {\rm diam}\, X )}
$$
where $C$ is the constant appearing in \equ(iib2)
and $f_\GI$ the monotonic function defined in
\equ(fGI) (definition \ref{non root}).

\end{theo}
\vv

\begin{rema}\label{perco2}
The theorem \ref{sup} implies that
the percolation probability $\theta_{p,q}(x_0\leftrightarrow\infty)$
is analytic in $p$ and is of the order $1- (1-p)^{\D}$ uniformly in $x_0$,
since $\theta_{p,q}(x_0\leftrightarrow\infty)= 1- \phi^{\rm f}_{p,q}(x_0)$.
In other words, the random cluster model on
percolative graphs
has a percolation probability
threshold $p_c$ strictly less than 1. On the other hand theorem \ref{sub} immediately implies
that  that $p_c>0$ in any bounded degree graph, and since any percolative graph is bounded degree,
we have immediately the corollary below, which can be considered as a generalization, for values
of $0<q<1$ and for percolative graphs, of theorem 4.2 in \cite{A} stated  for $\GI=\Z^d$ and $q\ge 1$.
\end{rema}

\begin{coro}
Let $\GI$ be an infinite graph and consider
the  random cluster model on $\GI$ with free or wired boundary conditions. Then, if $\GI\in\LL$,
for any $q>0$, the critical percolation probability defined in \equ(pc)
is such that  $p^{\x}_c(q)<1$, with $\x=0,1$.
\end{coro}
We remark that, due to the lack of validity of FKG inequalities, in the region $q<1$ we cannot conclude that
the percolation probability is monotonic increasing with $p$, so in principle in this region cannot be
excluded the possibility of more than one critical point.

\begin{rema}\label{decaypol}
The theorem \ref{sup}  also suggests that the fall-off rate  of the  finite connectivity functions at large
distances in general graphs in the highly supercritical phase may not necessarily be exponential, depending on the behavior of the function
$f_\GI$ defined in \equ(fGI). In particular, for graphs such that  $f_\GI(n)\approx C\ln n$ it seems reasonable to conjecture that
the finite connectivity
functions decay polynomially. We plan to prove such claim (searching for a lower bound on the finite connectivities)
in a future paper at least for $q=1$ (i.e. Bernoulli percolation) where calculations are much simpler.

\end{rema}

\\We now state the second theorem concerning the pressure.

\begin{theo}\label{presup}
Let $\GI=(\VU,\EE)\in \LL\cap\AA\cap\QQ^v\cap \QQ^e$, let $\{V_N\}_{N\in \mathbb{N}}$
be any
F\o lner sequence in $\VU$ such that
$V_N\nearrow\VU$, and let $(1-p)$ so small   that   $eA(1+\D^{R+1})\d_p\le 1$ where $\d_p$
is defined
in \equ(defel0).
Then
the pressure of Random Cluster Model on $\GI$, defined in \equ(press) exists
and can be written explicitly in term of an absolutely convergent
series which is analytic as a function of $p$, and  does not depend on $V_N$
and on $\x$.
\end{theo}

\subsection{Proof of theorem \ref{sup}.
Polymer expansion for the finite connectivity
functions}

\\In this section we will assume that $\GI=(\VU,\EE)$ is
percolative with maximum degree $\D$ , with cut-set constant $R$ and with
cut-set function $f_\GI$. We will
also assume that  $\{V_N\}$ is a
cut-set bounded sequence  in $\GI$ such that $V_N\nearrow \VU$.

\\The finite volume free and wired finite connectivity functions
for any $X\subset V_N-\partial^{\rm int}_vV_N$ can be written as

$$
\phi^{{\rm f},N}_{p,q,\x}(X)=
 {1\over \bar Z^\x_{N}(p,q)}\sum_{\om\in \O^\x_{\GI_{N}}:~\exists
g\in { A}_{\GI}:~ E_g\subset O(\om)\atop
X\subset V_g,~V_g\cap ~\partial^{\rm int}_v V_N=\emptyset}
\l^{|C(\om_N)|}q^{k_N^\x(\om)}
\Eq(connef5)
$$
\\where in this section
$$
\l={1-p\over p}
$$
and

$$
\bar Z^\x_{N}(p,q)=\sum_{\omega\in \Omega^\x_{\GI_N}}\l^{|C(\omega_N)|}
q^{k_{N}^\x(\om)}=p^{|\EE_N|}Z^\x_{{\GI_{N}}}(p,q)\Eq(bXi)
$$

\\We recall that the symbol $C(\om_N)$ denotes the set of closed edges
in $\EE_N$ once the configuration $\om \in \Omega^\x_{\GI_N}$ is given.

\vv
\def\DD{D}
\begin{defi}\label{dual}
A subset $S\subset \EE$ is  called a dual animal  if it is finite
and it is $R$-connected.
We say that two dual animals $S$ and $S'$ are {\it compatible} and we write
$S\sim S'$ if $S\cup S'$ is not a dual animal (i.e. $d_\GI(S,S')>R$).
We will denote by
$\E_{\GI}$ the set of all dual animals in $\EE$.  We will also denote by
$\E_{N}$ the set of dual animals in $\EE_N$.
\end{defi}

\vv

\\Observe that, since $\GI$ is assumed to be cut-set bounded, every fence in
$\GI$ is a dual animal.

\vv

\begin{defi}\label{contmin}
Let $S\subset \EE$ and let $\g\subset S$ be a fence with vertex interior
$V_\g$ and edge interior  $E_\g$.
We say that $\g$ is   minimal with respect to $S$ if there is no other fences
$\g'\subset S$ such that $\g'\cap\g\ne\0$
and $\g' \subset\g\cup E_\g$  (recall: $E_\g$ is the edge interior of $\g$). Note that a minimal fence $\g$
can contain in its interior
a fence $\g'$  such that $\g\cap\g'=\0$. Given $S\subset \EE$ we denote by
$ n_S$ the number of  fences which are minimal with respect to $S$.
\end{defi}

\begin{rema}
\\By the definition above and  by definition
\ref{fence}, if
$S\subset \EE$ is finite, then the number of finite connected component of
$(\VU, \EE-S)$ is exactly $n_S$.
\end{rema}

\\We will now give convenient expressions for $k_N^0(\om)$ and $k_N^1(\om)$.
Let us consider first the case $k_N^1(\om)$
which is the easier one.
If we are using wired boundary conditions, then $k_N^1(\om)$  is the
number of connected
components of $O(\om_N)$ plus the isolated vertices whose support is
contained in $V_N^{\rm int}$. The fences associated with any of
such components
is then totally contained in $\EE_N$. This means that
$$
k_N^1(\om)= n_{C(\om_N)}
\Eq(k1sc)
$$

\\Using now \equ(k1sc) the partition function $\bar Z^1_{{\GI_{N}}}(p,q)$
defined in \equ(bXi) can be rewritten as

$$
\bar Z^1_N(p,q)=\sum_{\omega\in \Omega^1_{\GI_N}}\l^{|C(\omega_N)|}
q^{k_{N}^1(\om)}=\sum_{\omega\in \Omega^1_{\GI_N}}\l^{|C(\omega_N)|}q^{n_{C(\om_N)}}
\Eq(bXi2)
$$

\\and

$$
\phi^{{\rm f},N}_{p,q,1}(X)=
 {1\over \bar Z^1_{\GI|_V}(p,q)}\sum_{\om\in \O^1_{\GI_{N}}:~\exists g\in { A}_{\GI}:~ E_g\in O(\om)\atop
X\subset V_g,~V_g\cap ~\partial^{\rm int}_v V_N=\emptyset}
\l^{|C(\omega_N)|}q^{n_{C(\om_N)}}
$$

\\The case $k_N^0(\om)$ is more involved. Observe first that the term in the partition function

$$
\bar Z^0_N(p,q)=\sum_{\omega\in \Omega^0_{\GI_N}}\l^{|C(\omega_N)|}
q^{k_{N}^0(\om)}
$$
corresponding  to the configuration
in which all bonds are open  is $q$ (since  $k^0_N(\om)=1$ in this case). For technical reasons is convenient
that this term is 1 (as it is in $\bar Z^0_N(p,q)$). So we define
$$
\hat Z^0_N(p,q)=\sum_{\omega\in \Omega^0_{\GI_N}}
\l^{|C(\omega_N)|}q^{k_{N}^0(\om)-1}\Eq(zhat)
$$
whence
$$
q\hat Z^0_N(p,q)=\bar Z^0_N(p,q)\Eq(zbarzhat)
$$
in such a way that $\hat Z^0_N(p,q)$ can be interpreted as a partition
function with term equal to 1
corresponding to the configuration in which all edges are open.

\\Now, by definition we can write

$$
\phi^{{\rm f},N}_{p,q,0}(X)=
 {1\over   \hat Z^0_N(p,q)}\sum_{\om\in \O^\x_{\GI_{N}}:~
\exists g\in { A}_{\GI}:~ E_g\in O(\om)\atop
X\subset V_g,~V_g\cap ~\partial^{\rm int}_v V_N=\emptyset}
\l^{|C(\omega_N)|}q^{k_{N}^0(\om)-1}
$$

We have now to write the explicit expression of $k_{N}^0(\om)$.
In this case we have to count the fences in the set $C(\om_N)\cup
\partial_eV_N\equiv\bar C(\om_N)$, and therefore we
allow fences $\bar\g$ such that
$\bar\g\cap\partial_eV_N\ne\emptyset$; in the latter case
the set $g\equiv\bar\g-\partial_eV_N$ will be called from now on {\it wall}.
Observe that since $V_N$ is a cut-set bounded sequence
(see definition \ref{seque}), then a wall in $\EE_N$
is $R$-connected, i.e. is a dual animal.

\\The number $k_{N}^0(\om)$
is then simply
$$k_{N}^0(\om)=n_{\bar C(\om_N)}
$$

\\Let us define for a given $S\in \E_N$
$$
\tilde n_{S} =\cases{ n_{S}& if $S\cup \partial_eV_N\notin\E$ \cr\cr
 n_{S\cup \partial_eV_N}-1& if $S\cup \partial_eV_N\in\E$ \cr\cr
}
\Eq(tilnen)
$$
and its activity $\r^\x(S)$  as follows

$$
\r^\x(S)= \cases{ \l^{|S|}q^{n_S}   &if $\x=1$\cr\cr
 \l^{|S|}q^{\tilde n_S}   &if $\x=0$
}
\Eq(acte)
$$
Defining
$$
\d_p=  \max\{ (|\l| q), |\l|  \}\Eq(defel)
$$

\\We have

$$
|\r^\x(S)|\le \d_p^{|S|}, \Eq(boundro)
$$

\\The reason why  we need to define for free boundary conditions the
quantity $\tilde n_S$ is the following: for a fixed dual animal
containing a wall, we can obtain a fence from the union of the
wall and the (closed) boundary in two different ways, while
we want to count the unit increasing of the number of connected components
of the configuration. This is the reason of the $-1$ in the definition of
$\tilde n_S$.

\\Furthermore, define the hard core pair
potential between two dual animals $S_i, S_j$ as
$$
U(S_i, S_j)=\cases{ +\infty &if $S_i\not\sim S_j$\cr\cr
0 &otherwise,}
\Eq(Ugtd)
$$

Use the shorthand notations

$$
\ES_n= (S_1, \dots,S_n)
~;
\quad{\r}^\x(\ES_n)\equiv\r^\x(S_1)\cdots\r^\x(S_n);
\quad
U(\ES_n)= \sum_{1\le i<j \le n}U(S_i, S_j)
$$

\\Then define the $\x$ dependent (for $\x=0,1$) polymer gas partition function as

$$
\Psi^\x_N(p,q)= 1+  \sum_{n\ge 1}{1\over n!}
\sum_{\ES_n\in (\E_{N})^n} \r^\x(\ES_n)
e^{-U(\ES_n)}
\Eq(Psi)$$
\\where  $(\E_N)^n$ is the $n$-times cartesian product of $\E_N$.
Note that, by construction
$$
\Psi^1_N(p,q)= \bar Z^1_N(p,q), ~~~~~ \Psi^0_N(p,q)= \hat Z^0_N(p,q) \Eq(psimu)
$$

\\and also

$$
\phi^{\rm f}_{p,q,\x}(X)=
{1\over \Psi^\x_N(p,q)}\sum_{n\ge 1}{1\over n!}
\sum_{\ES_n\in (\E_N)^n\atop \ES_n \bigodot X} \r^\x(\ES_n)
e^{-U(\ES_n)}\Eq(phisp)
$$

\\where condition $\ES_n \bigodot X$ on the sum above
means  that there must exist a fence
$\g\subset \cup_{i=1}^n S_i$  such that  $\g\bigodot X$ and
the set $\bar E_\g\cap[ \cup_{i=1}^n S_i]$ does not contains fences
$\g'$ such that $\g'\bigotimes X$ (here $\bar E_\g= \g\cup E_\g$).

\\We now rewrite the ratio \equ(phisp) (between two finite sums)
as a series. We follow and generalize the ideas developed in  \cite{BPS1}
and \cite{BPS2} for $\mathbb{Z}^d$. So we will define objects more general than dual animals
which will be called polymers.

\begin{defi}\label{polymer}
Let $X\subset \VU$ finite,
a set $P\subset \EE$  is called $X$-$R$-connected
if $P=\cup_{i=1}^k S_i$ with $k\ge 1$ and the following holds:
for all  $i=1,2,\dots ,k$   $S_i\in \E_\GI$; for all $i,j=1,2,\dots ,k$,
$S_i\sim S_j$ and
each $S_i$ contains a fence $\g_i$ such that $\g_i\bigodot Y$
for some non empty $Y\subset X$.
\end{defi}

\def\GG{{\mathcal G}}

\\We will denote by $\Pi^{X}$ the set of all $X$-$R$-connected sets
in $\EE$ and
by $\Pi^X_N$ the set of all $X$-$R$-connected sets in $\EE_N$.
We will also put $\E^{X}_\GI=\E_\GI\cup\Pi^X$
and $\E^{X}_N=\E_N\cup\Pi^X_N$.

\begin{defi}\label{Xpoly}
A set $P\in \E^{X}_\GI$ will be called a $X$-polymer
(or simply polymer when it is clear from the contest).
We will say that  two polymers $P_i\in  \E^{X}_\GI$ and $P_j\in  \E^{X}_\GI$  are
{\it compatible}, and we
write $P_i\approx P_j$, if
$P_i\cup P_j\notin  \E^{X}_\GI$; viceversa,
$P_i\in  \E^{X}_\GI$ and $P_j\in  \E^{X}_\GI$  are
{\it incompatible}, and we
write $P_i\not\approx P_j$, if
$P_i\cup P_j\in \E^{X}_\GI$.
\end{defi}
\def\P{\Pi}

\\Note that if $P \in \Pi^X$ and  $P'\in \Pi^X$ then necessarily
$P\not\approx P'$.

\\If $P\in \Pi^X$ and $P=\cup_{i=1}^k S_i$ with $k\ge 2$ we define the activity
of the polymer $P$ as $\r^\x(P)=\prod_{i=1}^k \r^\x(S_i)$. Define further
the hard core pair
potential between two polymers $P_i, P_j$ as
$$
\tilde U(P_i, P_j)=\cases{ +\infty &if $P_i\not\approx P_j$\cr\cr
0 &otherwise,}
\Eq(Ugtd2)
$$

\\Again, we use the shorthand notations

$$
\EP_n= (P_1, \dots,P_n)
~;\quad{\r}^\x(\EP_n)\equiv\r^\x(P_1)\cdots\r^\x(P_n);
\quad
\tilde U(\EP_n)= \sum_{1\le i<j \le n}\tilde U(P_i, P_j)
$$

\\Then, the r.h.s. of \equ(phisp) can be rewritten as
$$
\phi^{{\rm f},N}_{p,q,\x}(X)={1\over \Psi^\x_N(p,q)}\sum_{n\ge 1}{1\over n!}
\sum_{\EP_n\in (\E^X_{N})^n
\atop \exists! i\in \setn: ~P_i\bigodot X}
\r^\x(\EP_n)
e^{- \tilde U(\EP_n)}\Eq(phinf2)
$$

\\and the partition function  can be rewritten as

$$
\Psi^\x_N(p,q)=1+\sum_{n\ge 1}{1\over n!}
\sum_{\EP_n\in (\E^X_{N})^n}
\r^\x(\EP_n)
e^{- \tilde U(\EP_n)}
$$

\\Analogously as we did in section 4, we define, for $P\in \E^X$

$$
\Pi^{{\rm f},N}_{p,q, \x}(P)={\partial\over \partial \r^\x(P)}\ln\Big[ \Psi^\x_{{{N}}}(p,q)  \Big]
$$

\\Then, by construction
\def\gt{{\tilde \g}}
\def\Gt{{\tilde\Gamma}}
$$
\phi^{{\rm f}, N}_{p,q, \x}(X)=\sum_{P\in \E^X_N\atop
P\bigodot X}\r^\x(P)\Pi^{{\rm f},N}_{p,q, \x}(P)\Eq(PhiPiNs)
$$
Now, by standard cluster expansion it is  well known that
$$\ln \Psi^\x_{{{N}}}(p,q)  =
\sum_{n\ge 1}{1\over n!}\sum_{\EP_n \in (\E^X_N)^n}
\F^T (\EP_n)\r^\x(\EP_n)
\Eq(log2)
$$

\\where the Ursell coefficients
$\F^T (\EP_n)$ are given by
$$
\Phi^{T}(\EP_n)=
\cases{\sum\limits_{E\subset E(\EP_n)\atop(\setn,E)\in \GG_n}(-1)^{|E|}
&if $n\ge$ 2\cr\cr\cr
1&if $n=1$.}
\Eq(PhiTU2)
$$
where $E(\EP_n)=\{\{i,j\}\subset \setn: P_i\not\approx P_j\}$ and
$\GG_n$ denotes the set of all connected graphs with vertex set $\setn$.
So
$$
\Pi^{{\rm f},N}_{p,q, \x}(P)=\sum_{n\ge 0}{1\over n!}\sum_{\EP_n\in (\E^X_N)^n}
\F^T (P,\EP_n)\r^\x(\EP_n)
\Eq(PiN2)
$$

\\We also define
functions  on the whole $\GI$
(hence not depending on boundary conditions) as follows
$$
\Pi^{\rm f}_{p,q,}(P)=\sum_{n\ge 0}{1\over n!}\sum_{\EP_n\in (\E_\GI^X)^n}
\F^T (P,\EP_n)\r^\x(\EP_n)\Eq(Pi2)
$$

and
$$
\phi^{{\rm f}}_{p,q}(X)=
\sum_{P\in \E^X_\GI\atop
P\odot X}\r(P)\Pi^{\rm f}_{p,q,}(P)
\Eq(PhiPi2)
$$
which, as we will see,  represents an absolutely
convergent  expansion for $p$ near 1 for the infinite volume finite connectivity function.

\subsection{Proof of theorem \ref{sup}. Convergence of the finite connectivity functions}

\\As we did in section 4, we first prove an exponential bound on the activity $\r(R)$.

\begin{lem}\label{actysup}
Let $\GI$ be a cut-set bounded and bounded degree graph. Then
for any $n\ge 1$

$$
\sup_{e\in \EE}
\sum_{S\in {\E_\GI} \atop e\in S ,~|S|=n}1 + \sum_{P\in \Pi^X\atop|P|=n}1~~ \le ~~ A^n\Eq(boundd)
$$

\\where
$$
A=\Big[\max\{2C, 1\}\Big] \times
\Big[\D^{2R}\Big]\Eq(deltat)
$$
with $C$ being the constant appearing in \equ(iib2)

\end{lem}

\\{\bf Proof}. We start bounding the first term in r.h.s. of \equ(boundd)
i.e.
the number of dual animals of fixed cardinality  containing a fixed edge.
We recall that a dual animal is just a $R$-connected set of $\EE$. Thus recalling definition \ref{boundcut}
we have
$$
\sup_{e\in \EE}\sum_{S\in {\E_\GI}\atop e\in S ,~|S|=n}1\le \sup_{e\in \EE}
\sum_{S\subset \EE: \,\,S \,\,{\rm connected}\atop e\in S ,~|S|=Rn}1\le \D^{2Rn}\Eq(sumge)
$$

\\Concerning the second term in l.h.s. of \equ(boundd) this sum  is done only over Polymers
$P$ of the form $P=\cup_{i=1}^m S_i$ with $m\ge 1$ such that,
for all  $i=1,2,\dots ,m$:   $S_i\in \E_\GI$; for all $i,j=1,2,\dots ,m$,
$S_i\sim S_j$; and
each $S_i$ contains a fence $\g_i$ such that $\g_i\bigodot Y$ for some
$Y\subset X$. Hence

$$
\sum_{P\in \Pi^X:\, |P|=n}1
\le
\sum_{m=1}^n
~\sum_{k_1+\dots+k_m=n}~\prod_{i=1}^{m}
\Bigg[\sup_{x\in \VU}\sum_{S\in \E_\GI:~|S|=k_i
\atop \exists \g\subset S: \,\g\bigodot x}1\Bigg]
$$

\\Now, to bound the factor
$$
\sup_{x\in \VU}\sum_{S\in \E_\GI:~|S|=k_i
\atop \exists \g\subset S: \,\g\bigodot x}1
$$
we proceed as follows.
Since $\GI$ is connected and locally finite, for any $x\in \VU$   there exists a
geodesic ray $\r=(V_\r,E_\r)$ starting at $x$.
Then, since $S$ must contain a fence $\g$ such that $\g\bigodot x$, we
have, by proposition 2.2, that
$E_\r\cap \g\neq\emptyset$.
Let $e_{x}(\g)$ be the  first edge (in the natural
order of the ray)
in $E_\r$ which belongs to $\g$ and define

$$
r_{k_i}(x)=\{e\in E_\r: \exists \g\in \G_{\GI} ~\mbox{such that }~|\g|=k_i \mbox{ and }
e=e_{x}(\g)\}\Eq(rn)
$$

\\Hence
$$
\sup_{x\in \VU}\sum_{S\in \E_\GI:~|S|=k_i
\atop \exists \g\subset S: \,\g\bigodot x}1
= \sup_{x\in \VU}
 \sum_{e\in r_{k_i}(x)}
\sum_{{  S\in \E_\GI   \; |S|=k_i \atop
 \exists \g\subset S:\, \g\bigodot x}\atop e_x(\g)=e}1
\le \sup_{x\in \VU} |r_{k_i}(x)| \;\sup_{e\in \EE} \sum_{S\in {\E_\GI} \atop e\in S ,~|S|=k_i}1
$$
Now we observe that the interior $I_\g$ of $\g$ is a finite and
connected subset of $\VU$ and recalling the definition of the diameter
\equ(rayW) we have clearly that

$$
\sup_{x\in \VU} |r_n(x)|\le \sup_{x\in \VU}
\sup_{\g\in \G_\GI: \,\g\bigodot x\atop|\g|=n}{\rm diam}I_\g
$$
But, by \equ(iib) and \equ(iib2), we have immediately that  ${\rm diam}I_\g\le C^n$
so  we get that
$\sup_{x\in \VU} |r_n(x)|\le C^{n}$. Hence, recalling \equ(sumge)
$$
\sup_{x\in \VU}\sum_{S\in \E_\GI:~|S|=k_i
\atop \exists \g\subset S: \,\g\bigodot x}1\le
\Big[ C \D^{2R}\Big]^{k_i}
$$

\\so the second term

$$
\sum_{P\in \Pi^X\atop|P|=n}1\le \sum_{m=1}^n~
\sum_{k_1+\dots+k_m=n}~\prod_{i=1}^{m}
\Big[ C \D^{2R} \Big]^{k_i}
=\Big[ C \D^{2R} \Big]^{n}\sum_{m=1}^n~
\sum_{k_1+\dots+k_m=n} 1\leq
\Big[ 2C \D^{2R} \Big]^{n}
$$

$\Box$
\vv

\\We now prove the following lemma
\vv

\begin{lem}\label{decaysup}
For any $q>0$ the function  $\phi^{\rm f}_{p,q}(X)$ defined by \equ(PhiPiNs) is
analytic as a function of $p$  whenever
$eA(1+\D^{R+1})\varepsilon_p\le 1$ where $\varepsilon_p$ is defined in \equ(defel).
Moreover $\phi^{\rm f}_{p,q}(X)$ satisfies
the following bounds.
$$
|\phi^{\rm f}_{p,q}(X)|~\le~
(1+\D^{-R-1}) (Ae\d_p)^{f_\GI( {\rm diam}\, X )}
$$
where $A$ is the  constant defined in \equ(deltat) and $f_\GI$ is the monotonic function defined in
definition \ref{non root}.
Moreover, if $\{V_N\}_{N\in \N}$ is any cut-set bounded sequence
of subsets of $\VU$, then, for all $N\in \N$  the function  $\phi^{{\rm f},N}_{p,q,\x}(X)$
defined by \equ(connef5) is
analytic as a function of $p$  whenever
$eA(1+\D^{R+1})\varepsilon_p\le 1$.

\end{lem}

\\{\bf Proof}. We use here the Kotecky-Preiss condition \cite{KP}, which in this case can be checked easily. We stress that
our bounds are not optimal. So, the Kotecky-Preiss condition for
the polymer gas  with set of polmymers $P\in \E^X_\GI$ and with activity $\r^\x(P)$ states that
series \equ(log2), \equ(PiN2),
\equ(Pi2) converge  if  it is possible to find  $a>0$ such that for all polymers $P'\in \E^X_\GI$
$$
\sum_{P\in \E^X_\GI\atop P\not\approx P'} |\r^\x(P)|e^{a|P|}\le a|P'|\Eq(Kotpr)
$$
Recalling the estimate \equ(boundro), one can easily check that \equ(Kotpr)
becomes
$$
\sum_{n=1}^\infty(\d_p e^a)^n\sum_{P\in \E^X_\GI\, |P|=n\atop P\not\approx P'}\,1~
\le ~a|P'|\Eq(FPb)
$$
Now we have that

$$
\sum_{P\in {\E^X_\GI}:\, P\not\approx P'\atop |P|=n}1\leq
\sum_{S\in {\E_\GI}:\, |S|=n \atop  d_\GI(S,P')\le R}1+
\sum_{P\in \Pi^X:\, |P|=n}1\Eq(sumg)
$$

\\Now, let us define the edge set $B_R(P')=\{e\in \EE: d_\GI(e,P')\le R\}$,    then
$$
\sum_{S\in {\E_\GI}:\, |S|=n \atop  d_\GI(S,P')\le R}1\le |B_R(P')|\sup_{e\in \EE}
\sum_{S\in {\E_\GI} \atop e\in S ,~|S|=n}1
$$
We bound $|B_R(P')|$. Let $B^v_R(P')=\{v\in \VU: d_\GI(v,P')\le R\}$, then,
since $\GI$
has maximum degree $\D$ and since  each edge in $\EE$ is incident to two vertices in $\VU$
we have surely that
$$
|B_R(P')|\le{\D\over 2} B^v_R(P')\le {\D\over 2}\sum_{e\in P'} B^v_R(e)\le
{\D\over 2}|P'|\D^R\le
\D^{R+1} |P'|
$$
Whence the first term in r.h.s. of \equ(sumg) is bounded by
$$
\sum_{S\in {\E_\GI}:\, |S|=n \atop  d_\GI(S,P')\le R}1\le \D^{R+1} |P'| \;\;
\sup_{e\in \EE}
\sum_{S\in {\E_\GI} \atop e\in S ,~|S|=n}1
$$
Hence, by lemma \ref{actysup},  we have  that
$$
\sum_{{P\in\E^X_\GI}:\,P\not\approx P'\atop |P|=n}1\leq
\D^{R+1} |P'|\Bigg[\sup_{e\in \EE}
\sum_{S\in {\E_\GI} \atop e\in \g ,~|S|=n}1 + \sum_{P\in \Pi^X:\, |P|=n}1\Bigg]\le
\D^{R+1} |P'|A^n
\Eq(bnd)
$$
Hence \equ(FPb) becomes
$$
\sum_{n=1}^\infty(\d_p e^a)^n A^n~
\le ~{a\over \D^{R+1}}\Eq(FPc)
$$
choosing,  $a=1$ we get that the series \equ(Pi2) is absolutely converegent whenever
$$
\d_p\le {1\over eA(1+\D^{R+1})}
$$
and it is bounded by
$$
|\Pi^{\rm f}_{p,q,}(P)|\le e^{|P|}
$$
Whence, recalling \equ(PhiPi2)

$$
|\phi^{{\rm f}}_{p,q}(X)|\le
\sum_{P\in \E^X_\GI\atop
P\odot X}(e\d_p)^{|P|}
$$

\\Now let us find a lower bound for the number
$\min_{P\bigodot X}|P|$.

\\Let $U_X$ be a subset of $\VU$ definite as follows.
$U_X$ is connected, $X\subset U_X$
and
$|\partial_e U_X|$ is minimum, i.e if $U$ is another
connected subset of $\VU$ such that
$U\supset X$ then $|\partial_e U|\ge |\partial_e U_X|$.
Now since $P\bigodot X$ then by
construction
that
$|P|\ge |\partial_e U_X|$ since by definition  $P$ contains a
fence with vertex interior
containing $X$.
Now, since  $\partial_e U_X$ is a fence, then it is $R$-connected.
This means that

$$
|P|\ge |\partial_e U_X|\ge C\,f_\GI( {\rm diam}\, U_X )\ge C\,f_\GI( {\rm diam}\, X )
$$
So, using also \equ(boundd)
$$
|\phi^{{\rm f}}_{p,q}(X)|\le
\sum_{n\ge {1\over R}\, {\rm diam}\,X} (e\d_p)^{n}\sum_{P\in \E^X_\GI:\,|P|=n\atop
P\odot X}1\le \sum_{n\ge {1\over R}\, {\rm diam}\,X} (Ae\d_p)^{n}\le (1+\D^{-R-1}) (Ae\d_p)^{f_\GI( {\rm diam}\, X )}
$$
The proof of the second part of the lemma, i.e. the analitycity of  $\phi^{{\rm f},N}_{p,q,\x}(X)$
can be done in a similar way by observing that
$\phi^{{\rm f},N}_{p,q,\x}(X)$ admits the polymer representation \equ(phinf2) analogous to \equ(PhiPi2)
and $|\r^\xi(P)|\le \d_p$.
$\Box$

\vv
\\Now we prove the following lemma  which concludes the  proof of  theorem
\ref{sup}
\vv

\begin{lem}\label{limsup}
Let $\GI=(\VU, \EE)$ be a percolative graph and let
$\{V_N\}$  be any cut-set bounded sequence in $\VU$  such that
$V_N\nearrow\VU$ .
Then for any fixed $q>0$ and $p$ such that $eA(1+\D^{R+1})\d_p\le 1$, and $\x=0,1$

$$
\lim_{N\to \infty} \phi^{{\rm f},N}_{p,q,\x}(X) = \phi^{\rm f}_{p,q}(X)
$$
where $\phi^{\rm f}_{p,q}(X)$ is the function defined in \equ(PhiPi2).
\end{lem}

\\{\bf Proof}.
We will consider only the case $\x=0$,  which is the less trivial one.
$$
|\phi^{\rm f}_{p,q}(X)- \phi^{{\rm f},N}_{p,q,\x=0}(X) |\le
|\sum_{P\in \E^X_\GI\atop
P\odot X}\r(P)\Pi^{\rm f}_{p,q,}(P) -
\sum_{P\in \E^X_N\atop
P\bigodot X}\r^\x(P)\Pi^{{\rm f},N}_{p,q, \x}(P)|\le
$$

$$
=\Bigg|~\sum_{n\ge 1}{1\over (n-1)!}\sum_{\EP_n\in (\E^X_\GI)^n \atop
P_1\odot X}
\F^T (\EP_n)\r(\EP_n)
~
-~\sum_{n\ge 1}{1\over (n-1)!}\sum_{\EP_n\in (\E^X_N)^n \atop
P_1\odot X}
\F^T (\EP_n)\r^0(\EP_n)\Bigg|\le
$$
~

$$
\le ~\Bigg|\sum_{n\ge 1}{1\over (n-1)!}\Bigg\{\sum_{\EP_n\in (\E^X_\GI)^n:\,P_1\odot X \atop
\exists j\in \setn:~ P_j\not\subset \EE_N}
\!\!\!\!\F^T (\EP_n)\r(\EP_n)
~
+~\sum_{\EP_n\in (\E^X_N)^n: ~ P_1\odot X\atop
\exists j\in \setn:~  P_j\, {\rm contains\, a\,  wall\,}}
\!\!\!\!\!\!\!\!\F^T (\EP_n)\Big[\r(\EP_n)-\r^0(\EP_n)\Big]
\Bigg\}\Bigg|\le
$$
Using that $|\r(\EP_n)-\r^0(\EP_n)|\le 2\d_p^{\sum_{i=1}^n|P_i|}$,
due to the bound \equ(boundro), we get

$$
|\phi^{\rm f}_{p,q}(X)- \phi^{{\rm f},N}_{p,q,\x=0}(X) |~~
\le ~~\sum_{n\ge 1}{1\over (n-1)!}\sum_{\EP_n\in (\E^X_\GI)^n:\,P_1\odot X \atop
\exists j\in \setn:~ P_j\not\subset \EE_N}
\d_p^{\sum_{i=1}^n|P_i|}|\F^T (\EP_n)|
~~+
$$
$$
 ~~~~~~~
 ~~~~~~~~~+~~2\sum_{n\ge 1}{1\over (n-1)!}\sum_{\EP_n\in (\E^X_N)^n: ~ P_1\odot X\atop
\exists j\in \setn:~  P_j\, {\rm contains\, a\,  wall\,}}
\d_p^{\sum_{i=1}^n|P_i|}
|\F^T (\EP_n)|\Eq(2ser)
$$

\\Now, by lemma \ref{decaysup}, we already know that
for $eA(1+\D^{R+1})\d_p\le 1$ the two series in the left hand  side of inequality
\equ(2ser) are analytic in $\d_p$. Consider
the first term of the r.h.s. of \equ(2ser).
Let us split this term   in two series as follows
$$
\sum_{n\ge 1}{1\over (n-1)!}\sum_{\EP_n\in (\E^X_\GI)^n:\,P_1\odot X \atop
\exists j\in \setn:~ P_j\not\subset \EE_N}
\d_p^{\sum_{i=1}^n|P_i|}|\F^T (\EP_n)|=A_1+A_2
$$
with

$$
A_1=\sum_{n\ge 1}{1\over (n-1)!}\sum_{\EP_n\in (\E^X_\GI)^n:\,P_1\odot X \atop
\exists j\in \setn:~,~P_j\not\subset \EE_N,~P_i\neq P_j}
\d_p^{\sum_{i=1}^n|P_i|}|\F^T (\EP_n)|
$$
$$
A_2=\sum_{n\ge 1}{1\over (n-1)!}\sum_{\EP_n\in (\E^X_\GI)^n:\,P_1\odot X \atop
P_1\not\subset \EE_N}
\d_p^{\sum_{i=1}^n|P_i|}|\F^T (\EP_n)|
$$

Analyticity of $A_1$ as a function of $\d_p$ implies immediately that there exists a constant $C_1$ such that
${\rm C_1}\d_p<1$ and
$$
A_1\le ({\rm C_1} \d_p)^{n_0}
$$

\\where the lowest order   ${n_0}$ is

$$
n_0= \min_{{\EP_n\in ({\E^X_\GI})^n~\atop
G(\EP_n)\in \GG_n,~
P_1\bigodot X}\atop \exists j\in \setn:~ P_j\not\subset \EE_N,~P_i\neq P_j }
\{\sum_{i=1}^n|P_i|\}
$$
Here above the condition $G(\EP_n)\in \GG_n$ is due the presence the factor $\Phi^T(\EP_n)$.
It is easy to see that $n_0$ is at least
$$
n_0\ge
\min_{{\g\in\G_\GI,~\g\bigodot X,\atop ~S\in  \E_\GI,~ S
\not\subset \EE_N}\atop d_\GI(\g,S)\le R }
\{|\g|+|S|\}
$$
Now, by \equ(iib) and \equ(iib2), we have that $|\g|\ge \ln [{\rm diam}(I_\g)]$. So
$$
n_0\ge \ln\Bigg[
\min_{{\g\in\G_\GI,~\g\bigodot X,\atop ~S\in  \E_\GI,~ S
\not\subset \EE_N}\atop d_\GI(\g,S)\le R }
\{{\rm diam}(I_\g)+|S|\}\Bigg]\ge \ln\left[
\min_{x\in X } {1\over R} d_\GI(x,\partial V_N)\right]
$$
Thus, by  lemma \ref{graph}, the r.h.s. of inequality above is a divergent quantity
when $N\to\infty$. So we have shown that $A_1\to 0$ as $N\to \infty$. Concerning $A_2$ we have similarly
$$
A_2\le {\rm Const'} \d_p^{n'_0}
$$

\\where now
$$
n'_0= \min_{\EP_n\in ({\E^X_\GI})^n\atop
P_1\bigodot X,\,P_1\not\subset \EE_N} \{\sum_{i=1}^n|P_i|\}
~\ge ~\min_{P\bigodot X \atop P\not\subset \EE_N } \{|P|\}
$$
this can be easily bounded from below as
$$
n'_0
\ge \min_{S\in \EE_\GI: ~S\bigodot X \atop S\not\subset \EE_N } \{|\g|\}
$$

\\Similarly to the previous case, we have that the r.h.s. of the
inequality above diverges when $N\to\infty$.
$\Box$

\subsection{Proof of theorem \ref{presup}.}

\\In  this section, accordingly to the hypothesis of theorem \ref{presup},
 we will assume that
$\GI$ is amenable and quasi-transitive and that the sequence
$\{V_N\}_{N\in \N}$ is F\o lner.

\\By remark   \ref{mind}, if the pressure exists, it is independent of boundary conditions so we consider  here the case
$\x=1$ (wired boundary conditions)  which is  easier for $p$ near 1.

\\Recalling \equ(bXi), \equ(bXi2), \equ(Psi), \equ(psimu), the ``infinite volume''
pressure with wired boundary condition is given by
$$
\pi_{\mathbb{G}}(p,q)= -\lim_{N\to\infty}{|\EE_N|\over |V_N|}\ln p + \lim_{N\to\infty}{1\over |V_N|}
\ln \Psi^1_N(p,q)\Eq(presup1)
$$

\\We proved in proposition \ref{orbit}
the existence of the first limit in r.h.s. of \equ(presup1),
so to prove theorem \ref{presup}
we have to show the existence of the limit
$$
\lim_{N\to\infty}{1\over |V_N|}
\ln \Psi^1_N(p,q)\Eq(limpre)
$$
To do this we will use
the  simpler representation of $\ln \Psi^1_N(p,q)$ in terms of dual animals.
So recalling \equ(Psi) we can write
$$
\ln \Psi^1_N(p,q)= \sum_{n\ge 1}{1\over n!}
\sum_{\ES_n\in (\E_{N})^n}
\Phi^{T}(\ES_n) \r(\ES_n)
$$
where again we have used the short notation $\ES_n=(S_1,\dots,S_n)$ and  $\r(\ES_n)=\r(S_1)\dots \r(S_n)$.

\\We also define, for $e\in \EE$, the functions
$$
\varphi_\GI(e) = \sum_{n\ge 1}{1\over n!}
\sum_{\ES_n\in (\E_{\GI})^n \atop e\in S_1}
\Phi^{T}(\ES_n) {1\over |S_1|}\r(\ES_n)
$$
and
$$
F_N={1\over |V_N|}\sum_{e\in \EE_N}\varphi_\GI(e)\Eq(limm)
$$
It is easy to show, by checking the Kotecky-Preiss condition, that the three series above are  absolutely convergent
as soon as $eA(1+\D^{R+1})\varepsilon_p\le 1$
and hence $F_N$ and $\ln \Psi^1_N(p,q)$ are analytic in $\d_p$ and bounded at least by $C_1\d_p$ for some constant
$C_1$. Moreover, due to hypothesis that $\GI$ is edge quasi-transitive, $\varphi_\GI(e)$ takes
values in a finite set.

\\Consider now the limit
$$
\lim_{N\to \infty} F_N \doteq F_{\mathbb{G}}(q)\Eq(liF)
$$

By proposition \ref{orbit}  and via an argument completely analogous to that developed in
proposition 8 of \cite{PSG} adapted to edge quasi-transitive graphs,  the limit \equ(liF) exists. Note
that to prove the existence of the limit above one needs both vertex transitivity and edge transitivity.
Hence, as a limit of bounded analytic functions, $F_{\mathbb{G}}(q)$ is
is analytic in $p$ as long as $eA(1+\D^{R+1})\varepsilon_p\le 1$ and bounded by $C_1\d_p$.
This implies that the proof of the theorem is achieved if we show that
$$
\lim_{N\to\infty}{1\over |V_N|}
\ln \Psi^1_N(p,q)= F_{\mathbb{G}}(q)
$$
Observe that
$$
\log\Psi^1_N(p,q)- \sum_{e\in \EE_N}f_{\GI}(e)
=
\sum_{n=1}^\infty{1\over n!}\Bigg[
\sum_{\ES_n\in (\E_{N})^n}
\Phi^{T}(\ES_n) \r(\ES_n)-
 \sum_{e\in \EE_N}
\sum_{\ES_n\in (\E_{\GI})^n \atop e\in S_1}
\Phi^{T}(\ES_n) {1\over |S_1|}\r(\ES_n)\Bigg]
$$
Now note that
$$
\sum_{\ES_n\in (\E_{\GI})^n \atop e\in S_1}(\cdot )=
\sum_{\ES_n\in (\E_{N})^n \atop e\in S_1}
+
\sum_{{\ES_n\in (\E_{\GI})^n \atop e\in S_1}\atop \exists S_i: ~
S_i\not\subset \EE_N}
(\cdot )
$$
moreover

$$
\sum_{e\in \EE_N}\sum_{S_1\in \E_N\atop e\in S_1}(\cdot)=
\sum_{S_1\in \E_N}|S_1|(\cdot)~~~,~~~
\sum_{e\in \EE_N}\sum_{S_1\in \E_\GI\atop e\in S_1}(\cdot)=
\sum_{S_1\in \E_\GI\atop S_1\cap \EE_N\neq \0}|S_1\cap \EE_N|(\cdot)
$$

\\ hence, using also that ${|S_1\cap \EE_N|/ |S_1|}\le 1$ we get
$$
\Big|\log\Psi^1_N(p,q)- \sum_{e\in \EE_N}\varphi_{\GI}(e)\Big|
\le
\sum_{n=1}^\infty{1\over n!}
\sum_{{\ES_n\in [\E_\GI]^n\atop  S_1\cap \EE_N\neq \emptyset}
\atop\exists S_i: ~
S_i\not\subset \EE_N }
|\Phi^T(\ES_n)||\r(\ES_n)|
$$
Let  now choose $\ell>R \ln \D$ and define
$$
m_N^\ell ={1\over \ell}\ln \left[{|V_N|\over |\partial_e V_N|}\right]\Eq(Bns)
$$
Since by the hypothesis the sequence $ V_N$ is F\o lner,
then  $\lim_{N\to \infty}m_N^\ell=\infty$, for any  $\ell>0$.
We now can rewrite
$$
\sum_{{\ES_n\in [\E_\GI]^n\atop  S_1\cap \EE_N\neq \emptyset}
\atop\exists S_i: ~
S_i\not\subset \EE_N }(\cdot)
=
\sum_{{\ES_n\in [\E_\GI]^n\atop  S_1\cap \EE_N\neq \emptyset,~
|\ES_n|\ge m_N^\ell}
\atop\exists S_i: ~
S_i\not\subset \EE_N }(\cdot)
+
\sum_{{\ES_n\in [\E_\GI]^n\atop  S_1\cap \EE_N\neq \emptyset,~
|\ES_n|< m_N^\ell}
\atop\exists S_i: ~
S_i\not\subset \EE_N }(\cdot)
$$

\\Hence
$$
\left|\log\Psi^1_N(p,q)- \sum_{e\in \EE_N}\varphi_{\GI}(x)\right|\le
\sum_{n=1}^\infty{1\over n!}\sum_{{\ES_n\in [\E_\GI]^n\atop
S_1\cap \EE_N\neq \emptyset,~|\ES_n|\ge m_N^\ell}
\atop\exists S_i: ~
S_i\not\subset \EE_N }
 \left|\Phi^T(\ES_n)\r(\ES_n)\right|+
$$
$$
+
\sum_{n=1}^\infty{1\over n!}
\sum_{{\ES_n\in [\E_\GI]^n\atop  S_1\cap \EE_N\neq \emptyset,
~|\ES_n|< m_N^\ell}
\atop\exists S_i: ~S_i\not\subset \EE_N }
\left|\Phi^T(\ES_n)\r(\ES_n)\right|\Eq(d)
$$
The first sum can be bounded, for $2e\d<1$, by
$$
\sum_{n=1}^\infty{1\over n!}\sum_{{\ES_n\in [\E_\GI]^n\atop  S_1\cap \EE_N\neq \emptyset,~|\ES_n|\ge m_N^\ell}
\atop\exists S_i: ~
S_i\not\subset \EE_N }
 \left|\Phi^T(\ES_n)\r(\ES_n)\right|\le
{\rm Const.}|\EE_N|\d_p^{m_N^\ell}
$$
which, divided by $|V_N|$,
converge to zero as ${N}\to \infty$ because $|\EE_N|/|V_N|$ goes to a constant when $N\to \infty$ (see \equ(EnVn)) and
by hypothesis $m_N^\ell\to \infty$ as $N\to \infty$.

\\Concerning the second term in r.h.s. of \equ(d),
due to the factor $\Phi^T(\ES_n)$
the sets $S_i$ must be pair-wise
incompatible, which is to say $\cup_i S_i$ must be $R$-connected.
 Since $|\cup_i S_i|< \sum_i|S_i|< m_N^p$, from
the conditions $S_1\cap \EE_N\neq \0$ and  $S_i\not\subset \EE_N$, we conclude
that all polymers $S_i$ must lie in the set
$$
{\rm B}^e_{m_N^\ell}(\partial V_N)=\{e\in \EE   :
{1\over R}d_\GI(e, \partial_e V_N)\le m_N^\ell\}
$$
with cardinality bounded by
$$
|{\rm B}_{m_N^p}(\partial V_N)|\le |\partial_e V_N| \D^{R m_N^p+1}
$$
Hence we have that second sum in r.h.s. of \equ(d) is
bounded by
\vskip.2cm
$$
\sum_{n=1}^\infty{1\over n!}
\sum_{{\ES_n\in [\E_\GI]^n\atop  S_1\cap \EE_N\neq
\emptyset,~|\ES_n|< m_N^\ell}
\atop\exists S_i: ~S_i\not\subset \EE_N }
\left|\Phi^T(\ES_n)\r(\ES_n)\right|
\le{\rm Const'.}|\partial_e V_N| \D^{R m_N^\ell}\d
$$

\\ Thus  recalling definitions \equ(limm) and
\equ(Bns), we have
$$
\left|{1\over |V_N|}\log\Psi^1_N(p,q)-
{1\over |V_N|}\sum_{x\in V_N}\varphi_\GI(x)\right|=
\left|{1\over |V_N|}\log \Xi_{\GI|_{V_N}}-
F_{\mathbb{G}}(q)\right|
\le
$$
$$
\le {\rm Const.}{|\EE_N|\over |V_N|}\d_p^{m_N^\ell}+ {\rm Const'.} {|\partial_e V_N|\over| V_N|} \D^{R m_N^\ell}\d
$$
$$
\le {\rm Const.}  \left[{|\partial V_N|\over |V_N|}\right]^{|\ln\d_p)|\over \ell}
+
{\rm Const.} \d \left[{|\partial V_N|\over  |V_N|}\right]^{1-{R\ln\D\over \ell}}
$$
Since by hypothesis $|\partial V_N|/ |V_N|\to 0$ as $N\to \infty$, we conclude that the quantity above
is as small as we please for $N$ large enough. This ends the proof of the theorem. $\Box$

\vv
\\{\bf Acknowledgements}.
\v
\\ We thank an anonymous referee for his careful work of revision and for many useful comments, remarks and suggestions
which helped to improve the quality of the paper.
A.P. also thanks the brazilian agencies
CAPES,  CNPq and FAPEMIG for financial support.

\end{document}